\documentclass[msom,nonblindrev]{informs3}
\OneAndAHalfSpacedXI

\usepackage{makecell}
\usepackage{multirow}
\usepackage{natbib}
 \bibpunct[, ]{(}{)}{,}{a}{}{,}%
 \def\bibfont{\small}%

\usepackage[utf8]{inputenc} 
\usepackage[T1]{fontenc}    
\usepackage{hyperref}       
\usepackage{url}            
\usepackage{booktabs}       
\usepackage{amsfonts}       
\usepackage{nicefrac}       
\usepackage{microtype} 
\usepackage{enumerate}
\usepackage{comment} 
\newcommand{\N}{\mathbb{N}}

\newcommand{\I}{\mathbb{I}}
\newcommand{\R}{\mathbb{R}}
\newcommand{\prob}{\mathbb{P}}

\newcommand{\E}{\mathbb{E}}

\usepackage{bm}

\usepackage{amsmath,amssymb}
\usepackage{array}

\usepackage[textsize=tiny]{todonotes}

\newcommand{\target}{\textsc{T}}
\newcommand{\numslot}{L}

\newcommand{\loss}{\textsc{loss}}

\newcommand{\multlb}{\underline{\alpha}}

\newcommand{\alloc}{\mathcal{X}}
\newcommand{\outcome}{x}
\newcommand{\payment}{\mathcal{P}}
\newcommand{\pay}{p}
\newcommand{\auc}{\mathcal{A}}
\newcommand{\prof}{\bm{b}^{\top}}

\newcommand{\well}{W}
\newcommand{\opt}{\textsc{OPT}}
\newcommand{\feas}{\mathcal{F}}

\usepackage{algorithm}
\usepackage{algpseudocode}
\usepackage{amsmath,amssymb,amsfonts}
\usepackage{enumitem}
\usepackage{bm}
\usepackage{thmtools,thm-restate}
\usepackage{cleveref}
\usepackage{subcaption}
\usepackage{graphicx}
\usepackage{soul}
\usepackage[normalem]{ulem}

\newenvironment{itquote}
  {\begin{quote}\itshape}
  {\end{quote}\ignorespacesafterend}

\newcommand{\cond}[1]{\left. \right|}

\newtheorem{theorem}{Theorem}[section]
\newtheorem{corollary}[theorem]{Corollary}
\newtheorem{lemma}[theorem]{Lemma}
\newtheorem{remark}{Remark}[section]
\newtheorem{proposition}[theorem]{Proposition}
\newtheorem{definition}{Definition}[section]
\newtheorem{example}{Example}[section]

\begin{document}

\TITLE{Individual Welfare Guarantees in the Autobidding World with Machine-learned Advice}
\ARTICLEAUTHORS{%
\AUTHOR{Yuan Deng}
\AFF{Google Research, \EMAIL{dengyuan@google.com} \URL{}}
\AUTHOR{Negin Golrezaei}
\AFF{Sloan School of Management, Massachusetts Institute of Technology, \EMAIL{golrezae@mit.edu} \URL{}}
\AUTHOR{Patrick Jaillet}
\AFF{Department of Electrical Engineering and Computer Science, Massachusetts Institute of Technology, \EMAIL{jaillet@mit.edu} \URL{}}
\AUTHOR{Jason Cheuk Nam Liang}
\AFF{Operations Research Center, Massachusetts Institute of Technology, \EMAIL{jcnliang@mit.edu} \URL{}}
\AUTHOR{Vahab Mirrokni}
\AFF{Google Research, \EMAIL{mirrokni@google.com}\URL{}}
} 

\ABSTRACT{
Online advertising channels  have commonly focused on maximizing total advertiser value (or welfare) to enhance long-run retention and channel healthiness. Previous literature has studied auction design by incorporating machine learning predictions on advertiser values (also known as \textit{machine-learned advice}) through various forms to improve total welfare. Yet, such improvements could come at the cost of individual bidders' welfare and do not shed light on how particular advertiser bidding strategies impact welfare.

Motivated by this, we present an analysis on an individual bidder's welfare loss in the autobidding world for auctions with and without machine-learned advice, and also uncover how advertiser strategies relate to such losses. In particular, we demonstrate how ad platforms can utilize ML advice to improve welfare guarantee on the aggregate and individual bidder level by setting ML advice as personalized reserve prices when the platform consists of \textit{autobidders} who maximize value while respecting a return-on-ad spent (ROAS) constraint. Under parallel VCG auctions with such  ML advice-based reserves, we present a worst-case welfare lower-bound guarantee for an individual autobidder, and show that the lower-bound guarantee is positively correlated with ML advice quality as well the scale of bids induced by the autobidder's bidding strategies. Further, we prove an impossibility result showing that no truthful, and possibly randomized mechanism with anonymous allocations can achieve universally better individual welfare guarantees than VCG, in presence of personalized reserves based on ML-advice of  equal quality. Moreover, we extend our individual welfare guarantee results to generalized first price (GFP)  and generalized second price (GSP) auctions. Finally, we present numerical studies using semi-synthetic data derived from ad auction logs of a search ad platform to showcase improvements in individual welfare when setting personalized reserve prices with ML-advice.
}

\maketitle


\section{Introduction}
\label{sec:intro}

Online advertisers have access to a vast array of digital advertising channels, such as social media, web display, keyword search, etc., from which they can procure ad impressions and drive user traffic. One possible way for these channels to improve overall attractiveness and retention is to design appropriate ad auction mechanisms that enhance advertisers' total welfare, which reflects the aggregate advertiser-perceived value for procured ad impressions on the channel.

For instance, consider advertisers whose ad campaign objective is to maximize ad clicks that direct users to landing pages of their services or products, as described in \cite{ghose2009empirical}. These advertisers' perceived value of procured ad impressions is their click conversion rate, and thereby ad channels' welfare maximization goal translates into improving the aggregate realized click conversion among all participating advertisers.



Academic literature has developed various approaches to improve total welfare, one of which involves predicting advertiser values by applying machine learning tools to data on users' interactions with ads. In the instance where welfare corresponds to click conversion, channels use ML algorithms to produce predictions (i.e., ML advice) on click conversion rates for impressions. See \cite{richardson2007predicting,sodomka2013predictive,mcmahan2013ad} or \cite{yang2022click} for a comprehensive survey on click predictions.

Having obtained ML advice on advertiser values, recent works such 
as \cite{deng2021towards,balseiro2021robust,deng2022efficiency} motivate the approach to augment existing ad auctions by directly setting personalized reserve prices for advertisers using such ML advice, and show theoretical guarantees on total welfare improvement.

Nevertheless, these results have two important issues. First, the pursuit of improving total welfare may not necessarily guarantee all individual advertisers benefit equally and may come at the expense of certain individual advertisers' welfare. For instance, larger advertisers acquiring more impressions while smaller advertisers receive fewer impressions could potentially harm the businesses of smaller advertisers and compromise the overall health of the channel in the long term. Second, welfare improvement guarantees are presented in a price of anarchy (POA) fashion, which measures the worst-case outcome total welfare compared to the maximum achievable (or efficient) welfare. However, these POA bounds are independent of advertiser bidding strategies and thus do not shed light on how particular advertiser bidding strategies in ad auctions impact individual or total welfare.

In light of these insufficiencies of existing results, in this work, we address the following questions:

\begin{itquote}
Given an advertisers' bidding strategy to procure impressions in ad auctions, how can platforms characterize the potential welfare loss of  this individual advertiser?  How should ad channels utilize machine-learned advice that predict advertiser values to improve individual welfare?
\end{itquote}

We study a prototypical \textit{autobidding setting} where advertisers compete simultaneously in numerous multi-slot position auctions that are run in parallel, and aim to maximize total advertiser value under \textit{return-on-ad-spent (ROAS)} constraints that restrict total spend of a bidder to be less than her total acquired value across all auctions in an average sense; see similar setups in \citep{aggarwal2019autobidding,deng2021towards,balseiro2021robust,mehta2022auction}. On the other hand, ad platforms possess ML advice that predicts advertisers' real values with a certain degree of accuracy/quality. Under this setup, our main contributions and organization of the paper is described as followed:

\textbf{Strategy-dependent individual welfare guarantee metric for individual advertisers. }
In Section \ref{sec:model}, we present a novel individual welfare metric that measures the difference between two specific welfare outcomes of an individual advertiser: fixing a certain bidding strategy, the worst case welfare over all auction outcomes under which all bidders' ROAS constraints are satisfied; and the welfare this individual bidder would have obtained in the global welfare maximizing outcome. Our metric achieves two key goals: 1. it characterizes  individual welfare loss, and 2. allows platforms to uncover the relationship between advertiser strategies and individual welfare guaranties. 
Our proposed metric is the first of its kind to achieve these two goals.

\textbf{Individual welfare guarantees in VCG auctions by setting personalized reserves with ML advice. }
In Section \ref{sec:reserves}, we illustrate  through examples that setting ML advice as personalized reserves as in 
\citep{deng2021towards,balseiro2021robust,deng2022efficiency} 
surprisingly improves individual welfare guarantees under our aforementioned individual welfare metric. Then, in Section \ref{sec:VCG} where we consider parallel VCG auctions, and focus on an individual autobidder who adopts uniform bidding (Proposition \ref{prop:ubid}), we formally show that augmenting such auctions with ML-advice-based reserves allows us to present an individual welfare lower bound guarantee for this advertiser that increases in the advertiser's uniform bid multiplier, the quality of ML advice, and the relative market share of this advertiser compared to competitors (Theorem \ref{thm:fairnessbound}). 
Together with results in \citep{deng2021towards} stating ML-advice-based reserves can improve total welfare, we conclude that incorporating ML advice as personalized reserves achieves ``best of both worlds'' by simultaneously benefiting total and individual welfare. 

\textbf{Impossibility result: VCG yields the best individual welfare guarantees among a broad class of auctions. } In Section \ref{sec:imposs}, we show an impossibility result that says no allocation-anonymous, truthful, and possibly randomized auction format with ML advice of given quality can achieve a strictly better individual welfare guarantee than the VCG auction coupled with ML advice of the same quality; see Theorem \ref{thm:imposs}. In particular, for any allocation-anonymous, truthful, and possibly randomized auction, we construct a problem instance with personalized reserves based on ML-advice of given quality, and show that there must be at least 1 bidder whose welfare is at most the welfare lower bound guarantee we presented under VCG (i.e. Theorem \ref{thm:fairnessbound}).
    
\textbf{Extending individual welfare guarantee  to GSP and GFP.} We extend individual welfare guarantee results to GSP and GFP auctions, and show that a similar individual welfare lower bound guarantee for VCG continues to hold (Theorem \ref{thm:GFP:fairnessbound}). We compare these 
lower bound guarantees in GSP and GFP with that of VCG and identify conditions under which VCG outperforms (or underperforms) GSP/GFP in terms of our individual welfare metric with the same  ML advice quality. 

\textbf{Numerical results.} We present numerical studies using semi-synthetic data derived from auction logs of a  search ad platform to showcase individual welfare improvement via setting personalized reserve prices with ML-advice. We demonstrate that as ML-advice quality improves, more advertisers' welfare would approach what they would have obtained under the efficient outcome.

\subsection{Related Works}
\textbf{Autobidding and total welfare maximization.} The most relevant works to this paper are \cite{deng2021towards,balseiro2021robust,mehta2022auction}, where they consider the same autobidding setting (i.e. value-maximizers with ROAS constraints) as ours. \cite{deng2021towards,balseiro2021robust,mehta2022auction,deng2022efficiency} all present techniques to improve price-of-anarchy bounds for the total welfare of any feasible outcome in which all bidders' ROAS  constraints are satisfied: where \cite{deng2021towards} relies on additive boosts on bid values,
\cite{balseiro2021robust,deng2022efficiency} utilizes approximate reserve prices derived from ML-advice, and \cite{mehta2022auction} develops randomized allocation and payment rules. Our work distinguishes  itself from these works as we focus on welfare and individual welfare guarantees on the individual bidder level, and also sheds light on how autobidders' uniform bidding strategies affect individual welfare loss. We point out that our proof techniques  also differ from those in  \cite{deng2021towards,balseiro2021robust,mehta2022auction,deng2022efficiency} as our individual individual welfare guarantees require novel analyses on the value-expenditure tradeoffs individual bidders' would face when they are tempted to outbid others to acquire more value; see discussion in Section \ref{sec:VCG} for more details.

\textbf{Exploiting machine-learned advice.}
ML advice has been utilized in various applications to improve learning and decision making. For example, \cite{wang2020online} exploits ML advice to develop algorithms for the multi-shop ski-rental problem, \citep{lykouris2018competitive} adopts ML advice for the caching problem,  and \citep{indyk2022online} studies online page migration with ML advice. However, although many works in online advertising studied
predictive models for advertiser values, click through rates, etc (see e.g. \citep{richardson2007predicting,lee2012estimating,sodomka2013predictive}), the literature on applying such predictions (or more generally, ML advice) to the mechanism design problem has been scarce. See also \cite{correa2021secretary,golrezaei2022online} for works that exploit sample information (unstructured ML advice) in online decision-making.  One related work along this direction is \cite{munoz2017revenue}, which develops a theoretical framework to optimize reserve prices in a posted price mechanism by utilizing prediction inputs on bid values. In this work, we do not optimize for reserves, and motivate the simple approach of setting reserves using ML advice to improve individual advertiser welfare. Finally, we note that our work contributes to the area of exploiting ML advice to designing mechanisms for improving welfare guarantees for individual bidders.

For more related works mechanism design for constrained bidders, algorithmic bidding/learning under constraints, and reserve price optimization, we refer readers to Appendix \ref{app:extendlitrev} for an extended literature review.

\section{Preliminaries}
\label{sec:model}
We describe our model in the context of  sponsored search as in Section \ref{sec:intro}, but remark that all results and insights apply to general online advertising setups such as web display, e-commerce, social newsfeed, etc. Consider $N$ bidders (i.e. advertisers) participating in $M$ parallel position auctions $(\auc_{j})_{j\in[M]}$, where each auction $\auc_{j}$ is instantiated by a user keyword search query. An auction $\auc_{j}$ sells to bidders $\numslot_{j}\geq 1$ ad slots that are ordered by visual prominence, or equivalently the likelihood of the user viewing the slot, on the webpage, represented by click-through-rates (CTR) $1\geq \mu_{j}(1)\ge \mu_{j}(2)\ge \ldots \ge \mu_{j}(L_{j})\geq 0$, where $\mu_{j}(\ell)$ is the likelihood of the user of auction $j$ viewing slot $\ell \in [L_{j}]$ (see an intro to position auctions in e.g. \citep{lahaie2007sponsored,varian2007position,edelman2007internet}). A bidder $i\in [N]$ possess a private value-per-click equal to $v_{i,j}>0$ for auction $\auc_{j}$ that represents her perceived value conditioned on the user viewing her ad, so her attained utility for winning slot $\ell \in [L]$ is $\mu_{j}(\ell)\cdot v_{i,j}$.  

 In the following subsection \ref{subsec:auctionformat} we present an overview for position auction mechanisms; in subsection \ref{subsec:autobidder} we describe bidder objectives and actions; and finally in subsection \ref{subsec:outcome} we introduce definitions regarding bidder individual welfare guarantees. 
 
\subsection{Preliminaries for a single position auction} 
\label{subsec:auctionformat}
A (possibly randomized) position auction $\auc$ with $L\geq 1$ slots 
is characterized by a tuple $(\alloc,\payment, \bm{\mu})$, where $\alloc$ is an allocation rule, $\payment$ is a payment rule, and CTRs $\bm{\mu} = (\mu(\ell))_{\ell \in [\numslot]}\in [0,1]^{L}$ that satisfies  $1\ge \mu(1) \geq \mu(2)> \ldots \geq \mu(\numslot)\geq 0$. Let $N$ bidders with private value per-clicks $\bm{v} =(v_i)_{i\in [N]}$ participate in auction $\auc$ by submitting  a bid profile $\bm{b} =(b_i)_{i\in [N]} \in \R_{+}^{N}$, and we describe the payment and allocation rules as followed: 
 
The allocation rule $\alloc:\R_{+}^{N}\to \{0,1\}^{N\times L}$ maps bid profile $\bm{b} \in \R_{+}^{N}$ to an outcome $\bm{\outcome} = \alloc(\bm{b}) \in \{0,1\}^{N\times L}$ which may possibly be random. The entry
$\outcome_{i,\ell} = 1$ if bidder $i$ is allocated  slot $\ell \in [L]$, and 0 otherwise. Here, each slot $\ell$ is at most allocated to one bidder so $\sum_{i\in[N]}\outcome_{i,\ell} \leq 1$ for any $\ell$.  Further, under outcome $\bm{\outcome}\in \{0,1\}^{N\times L}$, bidder $i$ who has value $v_{i}$ attains a total welfare of $W_{i}(\bm{\outcome}) = v_{i}\sum_{\ell \in [L]}\mu(\ell) x_{i,\ell}$. That is, if bidder $i$ is allocated slot $\ell \in [L]$ (i.e., $x_{i, \ell} =1$), her welfare is 
$W_{i}(\bm{\outcome}) = \mu(\ell)v_{i}$. The payment rule $\payment:\R_{+}^{N}\rightarrow \R_{+}^{N}$  maps bids $\bm{b}$ to  payments $ \payment(\bm{b})\in \R_{+}^{N}$ where $\payment_{i}(\bm{b})$ is the payment of bidder $i$. In this work, we focus on the class of auctions that are \emph{ex-post individual rational} (IR), i.e. the payment for any bidder is less than her submitted bid, or mathematically $\payment_{i}(b_{i},\bm{b}_{-i})\leq b_{i}$ for any $\bm{b}_{-i} \in \R_{+}^{N-1}$. We note that the classic VCG, GSP and GFP auctions are ex-post IR. Further, we assume bidders who submit a 0 bid value will not be allocated any slots and incur no payment.



Having introduced general position auction  allocation and payment rules, in the following we define three special auction classes, namely truthful auction (Definition \ref{def:truthfulauctions}), allocation anonymous auctions (Definition \ref{def:anonalloc}), and personalized reserve augmented allocation anonymous auctions (Definition \ref{def:resaugment}).

\begin{definition}[Truthful auction]
\label{def:truthfulauctions} 
Consider position auction $\auc =(\alloc,\payment, \bm{\mu})$ where we recall $\alloc,\payment$ are possibly random allocation and payment rules, and $\bm{\mu}\in [0,1]^{L}$ are CTRs. Then we say the auction is truthful if for any bidder $i\in[N]$, her value $v_{i} \in \arg\max_{b\geq 0} \E\left[\well_{i}(\alloc(b,\bm{b}_{-i})) - \payment_{i}(b,\bm{b}_{-i})\right]$ for any competing bid profile $\bm{b}_{-i}$,  where the expectation is taken w.r.t. possible randomness in $(\alloc,\payment)$, and recall welfare $W_{i}(\bm{\outcome}) = v_{i}\sum_{\ell \in [L]}\mu(\ell) x_{i,\ell}$  with $\bm{x} = \alloc(b,\bm{b}_{-i})$.
\end{definition}
Note that the well-known VCG auctions is truthful. In truthful auctions it is a weakly dominant strategy for a bidder to bid her true value when her objective is to maximize quasi-linear utility. In the next Subsection \ref{subsec:autobidder} we study bidders whose objectives are not necessarily quasi-linear, so that truthful bidding is no longer weakly optimal in truthful auctions.

We next define allocation-anonymous auctions, in which if two bidders swap their bids, the probability of each bidder winning any slot will also be swapped, or in other words, the outcome of the position auction only depends on solicited bid values, and independent of bidders' identity.
\begin{definition}[Allocation anonymous auctions]
\label{def:anonalloc}
Consider position auction $\auc =(\alloc,\payment, \bm{\mu})$ and any permutation $\sigma:[N]\to [N]$ of $\{1\dots N\}$, as well as the permuted bid profile $\bm{b}' = (b_{\sigma(i)})_{i\in[N]}$. Let  $\bm{x} = \alloc(\bm{b})$, $\bm{x}'= \alloc(\bm{b}')$ be the (possibly random) outcomes under $\bm{b},\bm{b}'$, respectively. Then,  we say $\alloc$ is allocation anonymous if for any bidder $i\in[N]$ and slot $j\in [L]$, we have $\prob(x_{\sigma(i),j} = 1) = \prob(x_{i,j}' = 1)$. 
\end{definition}
 The classic VCG, GSP, and GFP are all allocation anonymous. The following presents an illustrative example of allocation-anonymity for GSP. 
\begin{example}[Example for allocation anonymous auctions]
Consider a single GSP auction with 2 slots and 3 bidders who submitted a bid profile $\bm{b} = (0.1,0.2,0.3)$. As GSP allocates slots by ranking bidders' submitted bids, the outcome under bid profile   $\bm{b}$  is $\bm{x} = \begin{pmatrix} 0, 0\\
0,1\\
1,0
\end{pmatrix}$.  Next, consider some permutation $\sigma$ that maps $\{1,2,3\}$ to $\{3,1,2\}$. That is, $\sigma(1) = 3$, $\sigma(2)=1$ and $\sigma(3) = 2$. Under this permutation, the corresponding permuted bid profile $\bm{b}' = (0.3,0.1,0.2)$, which  results in the outcome $\bm{x}' =  \begin{pmatrix} 1, 0\\
0,0\\
0,1
\end{pmatrix}$. Then, it is easy to check that  $ \prob(x_{\sigma(i),j} = 1) = \prob(x_{i,j}' = 1) = \begin{cases}
    1 & \text{if } (i,j) = (1,1) \text{ or } (3,2)\\
    0 & \text{otherwise} 
    \end{cases}$.
In particular,  because $\sigma(1) = 3$ we have $\prob(x_{3,1} = 1) = \prob(x_{1,1}' = 1) = 1$, and  because $\sigma(3) = 2$ we have $\prob(x_{2,2} = 1) = \prob(x_{3,2}' = 1) = 1$.
\end{example}

 

Finally, we describe augmenting allocation anonymous auctions with personalized reserves.
 \begin{definition}[Personalized-reserve augmented allocation anonymous auctions]
\label{def:resaugment}
Fix position auction $\auc =(\alloc,\payment, \bm{\mu})$, and some vector of personalized reserve prices $\bm{r} \in \R_+^{N}$ where  $r_{i}$ is the  reserve price for bidder $i \in [N]$. Then, the augmented auction is $\auc' =(\alloc',\payment', \bm{\mu})$ whose payment $\alloc'$ and allocation $\payment'$ are characterized via the following procedure for any  bid profile $\bm{b}\in \R_{+}^{N}$:
\begin{itemize}
    \item \textbf{$\alloc'$: } Define bid profile $\bm{b}' = (b_{i}\cdot \I\{b_{i}\geq r_{i}\})_{i\in [N]}$. Then $\alloc'(\bm{b}) =  \alloc(\bm{b}')$. \footnote{This allocation is known as an \textit{eager implementation} of personalized reserve prices, where any high-ranked slots are always allocated before a lower-rank slot gets allocated. There also exists a \textit{lazy implementations}, where we first rank all bids, and then allocate slots to each bidder following this ranking if the bidder clears her reserve. Note that the lazy implementation may leave ``holes'' in allocation, e.g. the first and third slots are allocated while leaving the second slot un-allocated. It will become clear later that all results in this work hold for both eager and lazy implementation of personalized reserve prices.} 
    \item \textbf{$\payment'$: } If $i\in[N]$ is not allocated a slot under outcome $\alloc'(\bm{b})$, $\payment_{i}'(\bm{b}) = 0$. Otherwise,  let $\ell_{i}\in[L]$ be the slot allocated to bidder $i$ under $\alloc'(\bm{b})$. Then,  $\payment_{i}'(\bm{b})  = \max\{ \payment_{i}(\bm{b}'), \mu(\ell_{i})\cdot r_{i}\}$. 
\end{itemize}
 \end{definition}
Recall that a 0 bid will always result in no allocation and 0 payment, so $\alloc'$ can be effectively viewed as excluding all bidders who do not clear their reserves, and implement the allocation rule $\alloc$ with respect to the remaining bidders. We remark in later sections, the personalized reserve prices relevant in this work (based on ML-advice) guarantee all bidders clear their reserves so that no bidders will be excluded from ranking. Finally, we refer readers to Example \ref{ex:augmentstandard} for an illustration of augmenting anonymous VCG, GSP, and GFP auctions with personalized reserves.

\subsection{Autobidders' objectives and bidding strategies}
\label{subsec:autobidder}
In this subsection we describe the scope for bidders' objectives as well as bidding strategies of interest. We recall the setup with $N$ bidders participating in $M$ parallel position auctions $(\auc_{j})_{j\in[M]}$, where $\bm{v}_{j}\in \R_{+}^{N}$ are bidders' values in auction $\auc_{j} = (\alloc_{j},\payment_{j},\bm{\mu}_{j})$  (see definitions in Subsection \ref{subsec:auctionformat}). We 
use the following notations for convenience:\\
 \renewcommand{\arraystretch}{2}
 \begin{center}
 \footnotesize
 \begin{tabular}[ht]{l cclcc}
 \hline
\hline
$\bm{b}_{i}\in \R_{+}^{M}$: bids submitted by bidder $i$ & &  & $\prof_{j}\in \R_{+}^{N}$: bid profile submitted to $\auc_{j}$\\
$\alloc_{j} (\prof_{j})\in \{0,1\}^{N\times \numslot_{j}}$: outcome of $\prof_{j}$ in $\auc_{j}$  & &  & $\payment_{j} (\prof_{j})\in \R_{+}^{N}$: payment vector of $\auc_{j}$\\
$\alloc_{i,\ell,j}(\prof_{j})\in  \{0,1\}$: indicator of bidder $i$ winning slot $\ell$ in $\auc_{j}$ & &  & $\payment_{i,j}(\prof_{j})\in \R_{+}$: payment of bidder $i$ in $\auc_{j}$ \\
 \hline
  \hline
\end{tabular}
\end{center}
\vspace{0.3cm}

Let $\alloc (\bm{b}):= (\alloc_j (\prof_{j}))_{j\in[M]}$.  Then, bidder $i$'s welfare in auction $\auc_{j}$ namely $W_{i, j}(\alloc_{j} (\prof_{j}))$, and her total welfare over all auctions namely $ W_i(\alloc(\bm{b}))$, are defined as 
 \begin{align}
\label{eq:defwelfare}
  W_i(\alloc(\bm{b})) :=\sum_{j\in [M]} W_{i,j}(\alloc_j (\prof_{j}))\quad \text{ and } \quad  W_{i,j}({\alloc}_{j}(\prof_{j}))  = \sum_{\ell=1}^{\numslot_{j}} \mu_{j}(\ell) \cdot v_{i,j}\cdot \alloc_{i,\ell,j}(\prof_{j})\,.
\end{align}


We study the setting where each bidder is subject to a \textit{return-on-ad-spent (ROAS)} constraint, which requires her total expenditure across all auctions to be less than her total acquired value.\footnote{A more general concept related to ROAS is \textit{return-on-investment (ROI)}, where each bidder $i$ has a target ROI ratio $\target_{i}$ such that her constraint in Eq. (\ref{eq:autobidderprob}) is instead written as $\well_{i}(\alloc(\bm{b}_{i},\bm{b}_{-i})) \geq \target_{i}\cdot \sum_{j\in[M]}p_{i,j}$; see e.g. \cite{golrezaei2021auction,golrezaei2021bidding}. In this paper, since we study worst-case  instances, we can scale all bidder $i$'s values by $\target_{i}$ so it is without loss of generality to consider ROAS constraints. } Mathematically, fix some competing bid profile $\bm{b}_{-i} \in \R_{+}^{(N-1)\times M}$, the ROAS constraint of bidder $i$ is 
\begin{align}
\label{eq:autobidderprob}
   \E\left[\well_{i}(\alloc(\bm{b}_{i},\bm{b}_{-i}))\right]\geq \E\left[\payment_{i}(\bm{b}_{i},\bm{b}_{-i})\right] 
    \quad \text{where } \payment_{i}(\bm{b}_{i},\bm{b}_{-i}) := \sum_{j\in[M]}\payment_{i,j}(\prof_{j})\,.
\end{align} 
Here, the expectation is taken w.r.t. possible randomness in 
the allocation and payment rules of auctions $(\auc_{j})_{j\in[M]}$. When allocation and payment for $(\auc_{j})_{j\in[M]}$ are deterministic (e.g. for VCG, GSP and GFP), we omit the expectation for simplicity. 

We call a bidder an \textit{autobidder} when she aims to maximize welfare $\E\left[\well_{i}(\alloc(\bm{b}_{i},\bm{b}_{-i}))\right]$ subject to the ROAS constraint in Eq. \eqref{eq:autobidderprob}. The following proposition states that an autobidder's optimal bidding strategy in truthful auctions, facing any competing bid profile, is \textit{uniform bidding}:
\begin{proposition}[Uniform bidding is optimal for autobidders in truthful auctions]
\label{prop:ubid}
Let all auctions $(\auc_{j})_{j\in[M]}$ be identical truthful auctions (see Definition \ref{def:truthfulauctions}), and bidder $i \in [N]$ is an autobidder who aims to maximize welfare $\E\left[\well_{i}(\alloc(\bm{b}_{i},\bm{b}_{-i}))\right]$ subject to the ROAS constraint in Eq. \eqref{eq:autobidderprob} for any fixed competing bids $\bm{b}_{-i} \in \R_{+}^{N-1}$. Then, there exists some constant uniform multiplier $\alpha_{i}^{*}\geq 1$ s.t. the uniform bidding profile $\alpha_{i}^{*} \bm{v}_{i}$ is $i$'s optimal strategy:
\begin{align}
\begin{aligned}
   \alpha_{i}^{*}\cdot \bm{v}_{i} ~\in~  \arg\max_{\bm{b}_{i}\in\R_{+}^{M}} ~& \E\left[\well_{i}(\alloc(\bm{b}_{i},\bm{b}_{-i}))\right] ~\text{s.t.}~ & 
    \E\left[\well_{i}(\alloc(\bm{b}_{i},\bm{b}_{-i}))\right] \geq \E\left[\payment_{i}(\bm{b}_{i},\bm{b}_{-i})\right],.
\end{aligned}
\end{align}
Further, adopting any uniform bid multiplier $\alpha_{i} < 1$ is weakly dominated by  truthful bidding, i.e.\\ $\E\left[\well_{i}(\alloc(\alpha_{i}\bm{v}_{i},\bm{b}_{-i}))\right] \leq\E\left[\well_{i}(\alloc(\bm{v}_{i},\bm{b}_{-i}))\right] $  for any $\bm{b}_{-i}\in \R_{+}^{N-1}$. 
\end{proposition}
This is a well-known result that has been proved and adopted in many related works such as \cite{aggarwal2019autobidding,deng2021towards,balseiro2021robust,mehta2022auction} and we will omit the proof here.  We remark that autobidding represents advertisers' conversion maximizing behavior while respecting constraints on spend. We note that our methodologies and insights in the paper can be extended to autobidders with a more general private costs
objective $\E\left[\well_{i}(\alloc(\bm{b}_{i},\bm{b}_{-i}))\right] - \rho \E\left[\payment_{i}(\bm{b}_{i},\bm{b}_{-i})\right] $ where $\rho\geq 0$ is some private cost, but for simplicity we assume in the rest of the paper $\rho = 0$. In light of this proposition, we will assume that all autobidders will adopt bid multiplier at least 1 in truthful auctions.

Finally, we conclude by introducing the notion of feasible bid profiles:
\begin{definition}[Feasible bid profiles]
\label{def:feas}
For a given set of parallel auctions $(\auc_{j})_{j\in[M]}$, we say that a bid profile $\bm{b}\in \R_{+}^{N\times M}$ is feasible if Eq. \eqref{eq:autobidderprob} holds for all bidders, and denote all feasible bid profiles as $\feas$. Further, fix a bidder $i$ and her bids $\bm{b}_{i}\in \R_{+}^{M}$. Then let $\feas_{-i}(\bm{b}_{i}) = \{\bm{b}_{-i}\in \R_{+}^{(N-1)\times M}: (\bm{b}_{i},\bm{b}_{-i}) \in \feas\}$.\footnote{Here, we exclude the trivial zero bid profiles in the sets $\feas$ and $\feas_{-i}(\bm{b}_{i})$.}
\end{definition}

In words, $\feas_{-i}(\bm{b}_{i})$ is all competing bid profiles for bidder $i$ that guarantee all bidders' ROAS constraints are satisfied. It is easy to see that in ex-post IR auctions, any bidder can satisfy her ROAS constraint by simply bidding truthfully, since a bidder's payment is always no greater than her submitted bid. This implies that the feasible set of bid profiles $\feas$ is never empty and contains the truthful bid profile.

\subsection{Efficient auction outcomes and individual welfare guarantees}
\label{subsec:outcome}
Let $\ell_{i,j}^{*}$ be the ranking of bidder $i$ in auction $\auc_{j}$ when ranked according to true values $\bm{v}_{j} \in \R_{+}^{N}$.  Then we call the outcome $\bm{\outcome}^{*} = (\bm{\outcome}_{j}^{*})_{j\in [M]}$ with  $\outcome_{i,\ell,j}^{*} = \I\{\ell = \ell_{i,j}^{*}\}$, the \textit{efficient outcome}. Note that $\bm{\outcome}^{*}$ yields the largest total welfare because allocation of slots in each auction follows ranking of bidder true values for that auction. Similar to our definition for welfare for any outcome in Eq. \eqref{eq:defwelfare}, let 
\begin{align}
\label{eq:optefficient}
    \opt_{i,j}= \mu_{j}(\ell_{i,j}^{*})\cdot v_{i,j}, \quad  \opt_{i}= \sum_{j\in[M]} \opt_{i,j}, \quad \text{and} \quad \opt =\sum_{i\in[N]} \opt_{i}
\end{align}
be the welfare of bidder $i$ in auction $j$, total welfare contribution of bidder $i$, and total welfare, respectively, under the  efficient outcome. Here we let $\mu_{j}(\ell) = 0$ for all $\ell > \numslot_{j}$.

 Despite truthful bidding is always feasible (see Definition \ref{def:feas} for feasible bid profiles and discussions on truthful bidding thereof), autobidders may not necessarily bid truthfully (even in truthful auctions) due to the presence of ROAS constraints, and instead adopt arbitrary strategies to optimize personal welfare which may deviate real auction outcomes from the efficient outcome. For certain individual bidders. Such deviations can potentially lead to significant welfare losses compared to the welfare she would have attained under the efficient one, whereas some other bidders may be significantly better off. These biases are not only largely unfavorable to bidders, but also to auction platforms as they may incentivize bidders to leave the platforms. 

 It is thus important for auction platforms to characterize to what extent the auctions  can provide individual welfare guarantee, and better understand how individual welfare relates to advertiser strategies. In the following Definition \ref{def:fairness}, we present a individual welfare metric that measures how much a bidder's realized welfare can fall short from the welfare she would have obtained in the efficient outcome if she adopts a certain strategy.
\begin{definition}[$\delta$-approximate]
\label{def:fairness}
Fix a bidder $i \in [N]$ and her bids $\bm{b}_{i}\in \R_{+}^{M}$. Then we say auctions $(\auc_{j})_{j\in[M]}$ are $\delta$-approximate for some $\delta\in [0,1]$ and any bidder $i\in [N]$ if
\begin{align}
    \min_{\bm{b}_{-i}\in \feas_{-i}(\bm{b}_{i})}\frac{\E\left[\well_{i}(\alloc(\bm{b})\right]}{\opt_{i}}~\geq~ \delta\,,
\end{align}
where $\feas_{-i}(\cdot)$ is defined in Definition \ref{def:feas}, the total welfare $\well_{i}$ under outcome $\alloc(\bm{b})$ is defined in Eq. \eqref{eq:defwelfare}, and the expectation is taken w.r.t. possible randomness in 
the allocation and payment rules of auctions $(\auc_{j})_{j\in[M]}$.
\end{definition}

The above individual welfare metric provides a quantitative answer to the following question: fixing a bid profile $\bm{b}_{i}$ for bidder $i$, among all outcomes induced by competing bid profiles $\bm{b}_{-i} \in \feas_{-i}(\bm{b}_{i})$ that ensure every bidders' ROAS constraint is satisfied (see Definition \ref{def:feas}),
what proportion of the welfare under the efficient outcome can be retained under the worst case outcome? We remark that under this interpretation, our individual welfare metric is reminiscent of the notion of price of anarchy (POA) which measures the worst-case total welfare achieved amongst all equilibrium compared to the optimal total welfare (see e.g. \cite{roughgarden2015intrinsic} for a detailed introduction on POA).

\section{Incorporating ML advice for bidder values as personalized reserve prices}
\label{sec:reserves}
With modern machine learning (ML) models and frameworks, 
online ad platforms can utilize available historical data (e.g. bid logs, keyword characteristics, user profiles, etc.) to produce predictions on autobidders' values (which we refer to as ML advice) ; see e.g. \cite{richardson2007predicting,sodomka2013predictive}. In this work, we specifically focus on ML advice that take the form of a \textit{lower-confidence bound} of true advertiser values. Our key approach to incorporate this type of ML advice in our autobidding setting, is via simply setting personalized reserve prices to be the lower confidence bound for each bidder's value. To motivate this approach, we start with an example.

\subsection{Motivating Example}\label{ex:motivate}

Consider 2 bidders competing in two (single-slot) second-price auctions (i.e. $\numslot_{1}=\numslot_{2}=1$) with corresponding CTRs $\mu_{1}(1) = \mu_{2}(1) = 1$. Bidders' values are indicated in the following table with some $v > 0$.
\vspace{0.5cm}
\renewcommand{\arraystretch}{1.5}
\begin{center}
\begin{tabular}{|c | c c| } 
 \hline
  & Auction 1 & Auction 2   \\
 \hline\hline
 bidder 1 & $v_{1,1} = v$ & $v_{1,2} = 0$ \\
 \hline
 bidder 2 & $v_{2,1} = \frac{v}{2}$  & $v_{2,2} =v$ \\
 \hline
\end{tabular}
\end{center}
\medskip

Suppose that both bidders are autobidders who adopt uniform bidding strategies (see Proposition \ref{prop:ubid}), and in particular,  suppose that  bidder $1$ sets her bid multiplier to be $\alpha_{1} = 1$. Then when her competitor bidder 2 sets a multiplier $\alpha_{2} > 2$, bidder 2 will win both auctions and acquire a total value/welfare of $v_{2,1} + v_{2,2}=\frac{3}{2}v$ while submitting a payment of $\alpha_{1}(v_{1,1} + v_{1,2}) = v$. In this case, bidder 2 satisfies her ROAS constraint and extracts all bidder 1's welfare, leaving her with no value. We also highlight that this bid mulitplier profile constitutes an equilibrium, \footnote{At an equilibrium bid multiplier profile, every bidder best responds to other bidders' bid multipliers while maintaining ROAS constraint satisfaction.} because bidder 1 cannot raise her bid multiplier to outbid bidder 2 for auction 1, since with $\alpha_{2} > 2$ bidder 1 would violate her ROAS constraint if she bids more than $\alpha_{2}v_{2,1} > v$.

Now suppose that  for each value $v_{i,j}$ ($i,j \in [2]$), the platform possesses a lower-confidence type of ML advice, namely  $(\underline{v}_{i,j})_{i,j\in[N]}$ such that $\beta v_{i,j}\leq\underline{v}_{i,j} < v_{i,j}$ for all $ v_{i,j}>0$ for some $\beta>\frac{1}{2}$, and sets personalized reserve price $r_{i,j} = \underline{v}_{i,j}$. If bidder 2 attempts to win both auctions by setting $\alpha_{2} > 2$, her payment will be at least  $\max\{\beta v_{2,1}, \alpha_{1} v_{1,1}\} +  \max\{\beta v_{2,2}, \alpha_{1} v_{1,2}\} = v + \beta v > \frac{3}{2}v$, violating her ROAS constraint. Therefore, via setting personalized reserves with ML advice, bidder 1's competitor is prohibited from outbidding her in auction 1, and hence safeguarding bidder 1's welfare.

\textbf{Key takeaway from Example \ref{ex:motivate}.} The main observation from the above example is that without reserve prices, bidder 2 acquires a large margin for her ROAS constraint by winning auction 2 where   payment required to win the auction is low. Therefore, she can raise her bid to outbid bidder 1 in auction 1 without violating her overall ROAS constraint. In other words, bidder 2 can compensate the high expenditure in auction 1 with her acquired value margin in auction 2.
By setting personalized reserve prices properly, the platform can increase bidder 2's payment in auction 2, which  in turn decreases the manipulative power of bidder 2. More generally, without reserve prices, bidders with large total values across auctions can overbid and consequently manipulate the outcome of certain auctions by compensating the incurring costs with acquired values from other auctions. Therefore, setting personalized reserve prices makes such overbidding behavior more costly, and thus reduces the overall manipulative power of bidders. 

\subsection{Setting personalized reserve prices using ML advice}
\label{subsec:approxreserves}
Here, we focus on  the following notion of \textit{approximate reserve prices} 
 with which we can reduce bidders' manipulative power as examplified in Example \ref{ex:motivate}.
 
\begin{definition}[$\beta$-accurate ML advice and approximate reserve prices]
\label{def:reserve}
Suppose there exists ML advice $(\underline{v}_{i,j})_{i,j\in[N]}$ in the form a lower-confidence bound. If 
$\underline{v}_{i,j}\in [\beta v_{i,j},v_{i,j})$ with some $\beta \in (0,1)$ for any bidder $i\in [N]$ and auction $j \in [M]$, we say the ML advice is $\beta$-accurate.  Further, if the platform sets $r_{i,j} = \underline{v}_{i,j}$, we say reserve prices $\bm{r}$ are $\beta$-approximate.\footnote{Note that any $\beta$-approximate reserve prices are also $\beta'$-approximate if $\beta'< \beta$. 
} 
\end{definition}


The gap between the lower bound $\beta v_{i,j}$ and the true value $v_{i,j}$ in Definition (\ref{def:reserve}) represents the inaccuracies of the platform's ML advice. In other words, $\beta$ can be perceived as a quality measure of the platform's ML advice for advertiser value, such that larger $\beta$ represents better advice quality. 

Further, ML-advice in online advertising settings generally concerns predicting advertiser values with historical conversion data  and produces confidence intervals of advertiser values (see e.g. \cite{shrestha2006machine,braga2007software,jiang2008estimating,dai2020coindice}). We remark that these confidence intervals can be viewed as a special case of the lower-confidence type of ML advice in Definition (\ref{def:reserve}): suppose the auctioneers  utilize some ML model to predict the true value $v_{i,j}$ of bidder $i$ in auction $j$, and produce a confidence interval $(\underline{v}_{i,j} , \Bar{v}_{i,j})\ni v_{i,j}$ with $\underline{v}_{i,j},\Bar{v}_{i,j} > 0$. The auctioneer can then choose personalized reserve $r_{i,j} = \underline{v}_{i,j}$, which is $\beta$-approximate for $\beta ={\underline{v}_{i,j}}/{\Bar{v}_{i,j}}\in (0,1)$ because $\beta v_{i,j} < \beta \Bar{v}_{i,j} = \underline{v}_{i,j} = r_{i,j} < v_{i,j}$.

Furthermore, in Definition \ref{def:reserve}, it is assumed that the ML advice $\underline v_{i,j}$ is a true lower bound on the bidder $i$'s value in auction $j$. This assumption can be relaxed by considering the ML advice that are accurate with high probability. 
Suppose the we possess some prediction $\Hat{v}_{i,j}$ for $v_{i,j}$ that satisfies $|\Hat{v}_{i,j} - v_{i,j}|< \eta$ with high probability (w.h.p) for some known $\eta$, then the confidence interval $(\Hat{v}_{i,j} - \eta, \Hat{v}_{i,j} + \eta)$ contains $v_{i,j}$ w.h.p. Them, the platform can set personalized reserve $r_{i,j} =\Hat{v}_{i,j} - \eta$. Note that with such personalized reserve prices derived from probabilistic ML-advice, all results in this paper remain valid w.h.p.

We conclude with a final remark regarding allocation-anonymous auctions (Definition \ref{def:anonalloc}).
\begin{remark}
\label{rmk:approxresanon}
We remark that implementing $\beta$-approximate personalized reserve prices in allocation-anonymous auctions does not impact anonymity, because $\beta < 1$ and thus all bidders clear there reserves. Therefore, the outcome with personalized reserves will be exactly the same as that without reserves; recall augmenting allocation-anonymous auctions with personalized reserves in Definition \ref{def:resaugment}. 
\end{remark}

\section{Individual welfare guarantees for VCG with ML advice}
\label{sec:VCG}
In the motivating Example \ref{ex:motivate}, we observe that ML advice and corresponding $\beta$-approximate reserves allow the parallel auctions to safeguard welfare for individual bidders by increasing payments and consequently limit the manipulative behavior of bidders who face significantly small competition in certain auctions. In this section, through the following Theorem \ref{thm:fairnessbound}, we formalize this intuition for the classic VCG auction, and present a quantitative measure for the relationship between overall individual welfare and ML advice when incorporated in the form of approximate reserves. 


 
\begin{theorem}[Individual welfare lower bound for VCGs with $\beta$-approximate reserves]
\label{thm:fairnessbound}
Consider the setting where $(\auc_{j})_{j\in [M]}$ are VCG auctions, and personalized reserve prices $\bm{r}$ are $\beta$-approximate as in Definition \ref{def:reserve}.
Fix an autobidder $i \in [K]$ who adopts bid multiplier $\alpha_{i} > 1$ (see Proposition \ref{prop:ubid}) so $\bm{b}_{i} = \alpha_{i} \bm{v}_{i}$. Then, the individual welfare guarantee in Definition \ref{def:fairness}) is bounded as:
\begin{align*}
  \min_{\bm{b}_{-i}\in \feas_{-i}(\alpha_{i}\bm{v}_{i})}  \frac{\well_{i}(\alloc(\bm{b}))}{\opt_{i}} ~\geq~ 1- \frac{1-\beta}{\alpha_{i}-1} \cdot \frac{\opt_{-i}}{\opt_{i}}\,,
\end{align*}
where $\feas_{-i}(\cdot)$ is defined in Definition \ref{def:feas}, and $\opt_{-i} = \sum_{j\neq i}\opt_{j}$.\footnote{We remark that the individual welfare lower bound in Theorem \ref{thm:fairnessbound} applies only to bidders whose welfare under the efficient outcome is nonnegative, i.e. $\opt_{i} > 0$.}
\end{theorem}
Details on implementation of  VCG with personalized reserve prices can be found in Definition \ref{def:resaugment} and Example \ref{ex:augmentstandard}. We defer our proof for Theorem \ref{thm:fairnessbound} to Section \ref{pf:thm:fairnessbound}, and here we provide some intuition for the individual welfare bound in the theorem.

In light of our interpretation for the individual welfare metric in Definition \ref{def:fairness}, Theorem \ref{thm:fairnessbound} states that at some bid multiplier $\alpha_{i}$, among all outcomes induced by competing bid profiles (possibly arbitrary non-uniform bidding profiles) that ensure every bidders' ROAS constraint is satisfied, in the worst case outcome,   bidder $i$ can retain at least a $1- \frac{1-\beta}{\alpha_{i}-1} \cdot \frac{\opt_{-i}}{\opt_{i}}$ portion of the welfare she would have obtained under the efficient outcome. We also remark that this individual welfare bound is very general since it does not impose any assumptions on competing bidders' bidding strategies and concerns any feasible general  bid profiles under which  bidders' ROAS constraint is satisfied. In other words, even if bidder $i$'s competing bidders optimize arbitrary objectives through complex bidding strategies, Theorem \ref{thm:fairnessbound} holds valid as long as the resulting bid profile is feasible.

The key message from Theorem \ref{thm:fairnessbound} is that with more accurate ML advice (i.e. larger $\beta$), auctions can set larger approximate reserves, and hence improve individual welfare guarantees for each individual bidder. We also provide some intuition for the term $ \frac{1-\beta}{\alpha_{i} -1}\frac{\opt_{-i}}{\opt_{i}}$ in the bound. Increasing $\beta$ (i.e.  increasing reserve prices via improving ML quality) or increasing the bid multiplier $\alpha_{i}$, raises the cost for competitors to outbid bidder $i$ in certain auctions, and hence makes it more difficult to cover her expenditures that arise from significant overbidding. This reduces competitors' manipulative power, and in turn improves the welfare guarantees for bidder $i$. Note that this aligns with the intuition we obtained in Example \ref{ex:motivate}. On the other hand, $\frac{\opt_{i}}{\opt_{-i}}$  can be perceived as the relative market share of bidder $i$ w.r.t. competing bidders. Our result shows that  with a small market share, the bidders become  vulnerable to manipulative behavior of others, resulting in low individual welfare guarantees.

The following Corollary \ref{cor:allbidderfair} presents a sufficient condition for the ML advice accuracy to achieve some predesignated level of individual welfare, whereas the following Theorem \ref{thm:VCGtight} states the individual welfare bound in Theorem \ref{thm:fairnessbound} is tight; see Appendix \ref{pf:thm:VCGtight} for corresponding proofs.
\begin{corollary}[ML advice accuracy level to achieve $\delta$-approximation of individual welfare]
\label{cor:allbidderfair}
Let $(\auc_{j})_{j\in[M]}$ be VCG auctions,
and assume all bidders adopt uniform bidding strategies with bid multipliers $(\alpha_{i})_{i\in [N]} \in (1,\infty)^{N}$, where we denote 
$\multlb = \min_{i\in [N]}\alpha_{i} > 1$. Then, with $\beta$ accurate  ML advice such that $\beta \geq 1- (1-\delta)\cdot (\multlb- 1) \cdot \min_{i\in [K]}\frac{\opt_{i}}{\opt_{-i}}$, the auctions $(\auc_{j})_{j\in[M]}$ are $\delta$-approximate for all bidders. 
\end{corollary}

\begin{theorem}[Matching individual welfare lower bound]
\label{thm:VCGtight}
For any $\beta \in (0,1)$, $\alpha > 1$, and $R \geq \frac{1-\beta}{\alpha-1}$, there exists values $\bm{v}\in \R_{+}^{N\times M}$ and $\beta$-approximate reserves $\bm{r}\in \R_{+}^{N\times M}$, such that there is a bidder $i$ with multiplier $\alpha_{i} = \alpha$ and relative market share $\frac{\opt_{i}}{\opt_{-i}} = R$, who has an individual welfare guarantee $\min_{\bm{b}_{-i}\in \feas_{-i}(\alpha_{i}\bm{v}_{i})}  \frac{\well_{i}(\alloc(\bm{b}))}{\opt_{i}} ~=~ 1- \frac{1-\beta}{\alpha_{i}  -1} \cdot \frac{\opt_{-i}}{\opt_{i}}$.
\end{theorem}

We conclude by comparing our individual welfare result in Theorem \ref{thm:fairnessbound} with related results in the literature: we point out that although our autobidding setup described in Section \ref{sec:model} and the notion of approximate reserves (Definition \ref{def:reserve}) are the same as those in \cite{deng2021towards,balseiro2021robust},
our analyses and proof techniques are different, primarily because we focus on the
welfare guarantees for individual bidders, where as \cite{deng2021towards,balseiro2021robust} investigates total welfare for all bidders. In particular, in our proof we fix a bidder $i$ and carefully analyze the amount of expenditure that could be covered by each competitor who outbids bidder $i$ in auctions where $i$ has the highest value, whereas the aforementioned related works takes an aggregate view to lower bound total welfare of all bidders; see more details in the next subsection \ref{pf:thm:fairnessbound}. Nevertheless, \cite{balseiro2021robust} shows that approximate reserves improve the total welfare of all bidders, and therefore along with Theorem \ref{thm:fairnessbound}, we can see that incorporating $\beta$-accurate ML advice as approximate reserves not only benefits total welfare, but also enhances individual welfare.

\subsection{Proof for Theorem \ref{thm:fairnessbound} and a tighter individual welfare guarantee}
\label{pf:thm:fairnessbound}
In this subsection, we first present the proof for the individual welfare lower bound in Theorem \ref{thm:fairnessbound}. From the proof, we further motivate a stronger individual welfare lower bound that depends on the total welfare of at most $\min\{M,N-1\}$ of bidder $i$'s competitors, instead of $\opt_{-i}$ which sums up the welfare of all $N-1$ of  bidder $i$'s competitors and can potentially be enormous due to large $N$. We will later remark that this strengthened bound, despite its improvement to the bound in Theorem \ref{thm:fairnessbound}, may be difficult to compute in practice.

\subsubsection{Proof for Theorem \ref{thm:fairnessbound}}
First, to prove Theorem \ref{thm:fairnessbound}, we rely on the definition of an advertisers' loss in welfare compared to  her  welfare contribution under the efficient outcome, formally defined as followed:
\begin{definition}[Welfare loss w.r.t. efficient outcome]
\label{def:welfareloss} 
For any bidder $i\in [N]$ and outcome $\bm{\outcome} = (\bm{\outcome}_{j}\in \{0,1\}^{N\times\numslot_{j}})_{j\in [M]}$, let $\mathcal{L}_{i}(\bm{\outcome}) = \{j\in [M]: \well_{i,j}(\bm{\outcome})< \opt_{i,j}\}$ be the set of auctions in which bidder $i$'s acquired welfare is less than that of her welfare under the efficient outcome. Then, we define the welfare loss of bidder $i$ under outcome $\bm{\outcome}$ w.r.t. the efficient outcome $\bm{\outcome}^{*}$ as:
\begin{align}
\label{eq:loss}
\loss_{i}(\bm{\outcome}) = \sum_{j \in \mathcal{L}_{i}(\bm{\outcome})}\left(\opt_{i,j}-\well_{i,j}(\bm{\outcome})\right)\,.
\end{align}
\end{definition}
\begin{remark}
\label{rem:rewritebadauctions}
For any outcome $\bm{\outcome}$, let $\ell_{i,j}$ be the position (i.e. ranking) of bidder $i$ in auction $j$, and recall that $\ell_{i,j}^{*}$ is the position of bidder $i$ in auction $j$ under the efficient outcome $\bm{\outcome}^{*}$. Then, the set $\mathcal{L}_{i}(\bm{\outcome}) = \{j\in [M]: \well_{i,j}(\bm{\outcome}_{j})< \opt_{i,j}\}$ (where $\well_{i,j}(\bm{\outcome}_{j})$ is bidder $i$'s welfare in $\auc_{j}$ as defined in Eq.\eqref{eq:defwelfare}) can also be interpreted as the set of auctions where bidder $i$'s ranking under $\bm{\outcome}$ is lower than her ranking under $\bm{\outcome}^{*}$, or in other words the set of auctions that incur a welfare loss w.r.t. $\bm{\outcome}^{*}$. Hence we can also rewrite $\mathcal{L}_{i}(\bm{\outcome}) = \{j\in[M]: \ell_{i,j} > \ell_{i,j}^{*}\}$.
\end{remark}
The following proposition connects the notion of welfare loss (as in Definition \ref{def:welfareloss}) and individual welfare (as in Definition \ref{def:fairness}) by showing  an upper bound on  welfare loss can be directly translated into a welfare lower bound that corresponds to our individual welfare guarantee. 
\begin{proposition}[Translating loss to individual welfare guarantee]
\label{prop:boundloss}
Assume for bidder $i \in [N]$ and outcome $\bm{\outcome}=(\bm{\outcome}_{j}\in \{0,1\}^{N\times\numslot_{j}})_{j\in [M]}$ we have $\loss_{i}(\bm{\outcome})\leq B$ for some $B > 0$. Then, $\frac{\well_{i}(\bm{\outcome})}{\opt_{i}}\geq 1- \frac{B}{\opt_{i}}$.
\end{proposition}

The proof of this proposition is presented in  Section \ref{pf:prop:boundloss}. Now, in light of this proposition, we proceed to prove Theorem \ref{thm:fairnessbound} by bounding bidder $i$'s welfare loss for auctions where she obtains a slot that is lower in position than what she would have obtained under the efficient outcome. 

\textit{Proof of Theorem \ref{thm:fairnessbound}.}
Fix any feasible competing bid profile $\bm{b}_{-i} \in\feas_{-i}(\alpha_{i}\bm{v}_{i})$ under which every bidders' ROAS constraint is satisfied; see Definition \ref{def:feas}. Denote the corresponding outcome as $\bm{\outcome} = \alloc(\bm{b})$, and $\ell_{k,j}$, $\ell_{k,j}^{*}$ to be the position of any bidder $k\in[N]$ in auction $j\in [M]$ under outcome $\bm{\outcome}$ and the efficient outcome, respectively. 

Consider any auction $j\in \mathcal{L}_{i}(\bm{\outcome}) =\{j\in[M]: \ell_{i,j} > \ell_{i,j}^{*}\} $ (see Remark \ref{rem:rewritebadauctions}), i.e. in auction $\auc_{j}$, bidder $i$ acquires a position (under $\bm{\outcome}$) bellow her position in the efficient outcome $\bm{\outcome}^{*}$. This implies there must exist competing bidders in auction $\auc_{j}$ whose values are smaller than that of bidder $i$'s, but obtains a better position, making bidder $i$ lose welfare. Motivated by this, we let $\mathcal{B}_{i}(k;\bm{x})$ denote the set of all auctions in which bidder $k$'s value is lower than $i$'s  but acquires a better position than $i$: 
\begin{align}
\label{eq:covering0}
    \mathcal{B}_{i}(k;\bm{x})= \left\{j\in [M]: \opt_{i,j} > 0,~ v_{k,j} < v_{i,j} \text{ and } \ell_{k,j} \leq \ell_{i,j}^{*} <  \ell_{i,j}\right\}
\end{align}
where we recall $\opt_{i,j} $ is the welfare of bidder $i$ in auction $j$ under the efficient outcome. Further, we can find a collection of $i$'s competitors whose $\mathcal{B}_{i}(~\cdot~;\bm{x})$ ``covers'' all auctions $\mathcal{L}_{i}(\bm{x})$ in which $i$ loses welfare. We call this collection of competitors a covering, and formally define the collection of all coverings, called $\mathcal{C}_{i}(\bm{\outcome})$, as followed:
\begin{align}
\label{eq:covering}
\begin{aligned}
  \mathcal{C}_{i}(\bm{\outcome}) = \left\{\mathcal{C} \subseteq [N]/\{i\}: (\mathcal{B}_{i}(k;\bm{x}))_{k\in \mathcal{C}} \text{ is a maximal set cover of }\mathcal{L}_{i}(\bm{\outcome})\right\}\,.
    \end{aligned}
\end{align}
Here, for any set $\mathcal{S}$, we say $\mathcal{S}_{1}\dots \mathcal{S}_{n}$ a maximal set cover of $\mathcal{S}$ if $\mathcal{S}\subseteq \bigcup_{n'\in[n]}\mathcal{S}_{n'}$ but $\mathcal{S}\subsetneq \bigcup_{n'\in[n]}\mathcal{S}_{n'} / \mathcal{S}_{n''} $ for any $n''\in [n]$.
 In words, $\mathcal{B}_{i}(k;\bm{\outcome})$ is the set of auctions in which bidder $k$ has a smaller value than bidder $i$ but acquires a better position, and any $\mathcal{C}\in \mathcal{C}_{i}(\bm{\outcome})$ is a subset of $i$'s competitors who are responsible for all welfare losses of bidder $i$ in auctions of $\mathcal{L}_{i}(\bm{\outcome})$. 

Fix any covering $\mathcal{C}\in \mathcal{C}_{i}(\bm{\outcome})$, and some bidder $k \in \mathcal{C}$. 
We first state the following inequality that bounds the welfare loss of bidder $i$ caused by competitor $k\in \mathcal{C}$ in the covering (we will prove this inequality later).
\begin{align}
\label{eq:welfarelossdue2k}
    \sum_{j\in \mathcal{B}_{i}(k;\bm{b})}\left(\mu(\ell_{i,j}^{*}) - \mu(\ell_{i,j})\right) v_{i,j}\leq \frac{1-\beta}{\alpha_{i}-\beta} \sum_{j\in[M]}  \mu(\ell_{{k},j})v_{{k},j} 
\end{align}
Summing the above over all competitors $k \in \mathcal{C}$, we have
\begin{align}
\label{eq:vcgboundloss}
\begin{aligned}
  \loss_{i}(\bm{\outcome}) ~=~& \sum_{j \in \mathcal{L}_{i}(\bm{\outcome})}\left(\mu(\ell_{i,j}^{*}) - \mu(\ell_{i,j})\right) v_{i,j}
  ~\overset{(a)}{\leq}~ \sum_{k \in \mathcal{C}}\sum_{j\in \mathcal{B}_{i}(k;\bm{b})} \left(\mu(\ell_{i,j}^{*}) - \mu(\ell_{i,j})\right) v_{i,j}\\
 ~\overset{(b)}{\leq}~& \frac{1-\beta}{\alpha_{i}-\beta} \sum_{k\in \mathcal{C}}\sum_{j\in[M]}  \mu(\ell_{{k},j})v_{{k},j}  
  ~=~
  \frac{1-\beta}{\alpha_{i}-\beta}
\sum_{k \in \mathcal{C}}\well_{k}(\bm{\outcome})\\
   ~\leq~&  \frac{1-\beta}{\alpha_{i}-
   \beta}\well_{-i}(\bm{\outcome})  \\
  ~\overset{(c)}{\leq}~&  \frac{1-\beta}{\alpha_{i}-\beta}\left(\opt_{-i}+\loss_{i}(\bm{\outcome})\right) \\
  ~ \Longrightarrow~  \loss_{i}(\bm{\outcome}) ~\leq~& \frac{1-\beta}{\alpha_{i}-1}\opt_{-i}\,.
  \end{aligned}
\end{align}
Here, in (a) we used the fact that
$\mathcal{L}_{i}(\bm{\outcome})\subseteq \bigcup_{k\in \mathcal{C}}\mathcal{B}_{i}(k;\bm{\outcome})$ (see  Eq. \eqref{eq:covering}); in (b) we applied Eq. \eqref{eq:welfarelossdue2k};
(c) follows from $\opt\geq \sum_{i \in [N]}\well_{i}(\bm{\outcome})$ where $\opt$ is the total efficient welfare and $\sum_{i \in [N]}\well_{i}(\bm{\outcome})$ is the total welfare under outcome $\bm{\outcome}$, so further
\begin{align}
\begin{aligned}
\label{eq:VCG:well2opt-ibound}
    \opt_{-i} ~\geq~& \well_{-i}(\bm{\outcome}) + \well_{i}(\bm{\outcome})  -\opt_{i} \\
    ~=~& \well_{-i}(\bm{\outcome}) + \sum_{j\in \mathcal{L}_{i}(\bm{\outcome})}\left(\well_{i,j}(\bm{\outcome}) -\opt_{i,j}\right) +  \sum_{j\in [M]/ \mathcal{L}_{i}(\bm{\outcome})}\left(\well_{i,j}(\bm{\outcome}) -\opt_{i,j}\right)  \\
    ~\overset{(e)}{\geq}~& \well_{-i}(\bm{\outcome}) + \sum_{j\in \mathcal{L}_{i}(\bm{\outcome})}\left(\well_{i,j}(\bm{\outcome}) -\opt_{i,j}\right) \\
    ~=~& \well_{-i}(\bm{\outcome}) - \loss_{i}(\bm{\outcome})\,.
    \end{aligned}
\end{align}
where in (e) we used the fact that $\well_{i,j}(\bm{\outcome}) \geq \opt_{i,j}$ in any auction $j\in [M]/ \mathcal{L}_{i}(\bm{\outcome})$. Finally, applying Proposition \ref{prop:boundloss} w.r.t. upper bound of $\loss_{i}(\bm{\outcome})$, and noting that the feasible competing bid profile is arbitrary, we obtain the desired welfare guarantee lower bound.

Now, it remains to prove Eq. \eqref{eq:welfarelossdue2k} that bounds the welfare loss of bidder $i$ caused by competitor $k\in \mathcal{C}$ in the covering. Denote $\pay_{k,j}$ as the payment of bidder $k$, and
$\Hat{b}_{\ell,j}$ as the $\ell$th largest bid in any auction $j\in[M]$.  Then in some auction $j\in \mathcal{B}_{i}(k;\bm{b})$ recall from Eqs. \eqref{eq:covering0} and \eqref{eq:covering} that $v_{k,j}<v_{i,j}$ but $ \ell_{k,j} \leq \ell_{i,j}^{*} <  \ell_{i,j}$. Thus bidder $k$'s payment is lower bounded as 
\begin{align}
\begin{aligned}
\label{eq:VCG:badbidder1}
   \text{ For } j \in \mathcal{B}_{i}(k;\bm{b}), \quad   p_{{k},j}~\overset{(a)}{\geq}~& \sum_{\ell = \ell_{{k},j}}^{\numslot_{j}}\left(\mu(\ell) - \mu(\ell+1)\right)\Hat{b}_{\ell+1,j}\\
    ~=~& \sum_{\ell = \ell_{{k},j}}^{\ell_{i,j}^{*}-1}\left(\mu(\ell) - \mu(\ell+1)\right)\Hat{b}_{\ell+1,j}+\sum_{\ell = \ell_{i,j}^{*}}^{\ell_{i,j}-1}\left(\mu(\ell) - \mu(\ell+1)\right)\Hat{b}_{\ell+1,j} +  p_{i,j}\\
     ~\overset{(b)}{\geq}~&  \left(\mu(\ell_{{k},j}) - \mu(\ell_{i,j}^{*})\right)v_{i,j} + \alpha_{i}\left(\mu(\ell_{i,j}^{*}) - \mu(\ell_{i,j})\right) v_{i,j} + \beta \cdot  \mu(\ell_{i,j})v_{i,j}\\
      ~=~&  \mu(\ell_{{k},j}) v_{i,j} + \left(\alpha_{i}-1\right)\left(\mu(\ell_{i,j}^{*}) - \mu(\ell_{i,j})\right) v_{i,j} - (1- \beta) \cdot  \mu(\ell_{i,j})v_{i,j}\,.
    \end{aligned}
\end{align}
Here , (a) follows from the VCG payment rule (see Example \ref{ex:augmentstandard}); (b) follows from the fact that bidder $i$'s ranking is $\ell_{i,j}$, so any bidder who is ranked before 
position $\ell_{i,j}$ submits a bid greater than bidder $i$'s bid $b_{i,j} = \alpha_{i}v_{i,j}$, i.e.
$\Hat{b}_{\ell,j} \geq b_{i,j} = \alpha_{i}v_{i,j} > v_{i,j}$ for any $\ell\leq \ell_{i,j}$. 

On the other hand, we have
\begin{align*}
      & \sum_{j\in \mathcal{B}_{i}(k;\bm{b})}p_{{k},j} + \sum_{j\notin \mathcal{B}_{i}(k;\bm{b})}  p_{{k},j} \leq \sum_{j\in \mathcal{B}_{i}(k;\bm{b})}\mu(\ell_{{k},j})v_{{k},j} + \sum_{j\notin \mathcal{B}_{i}(k;\bm{b})}  \mu(\ell_{{k},j})v_{{k},j}\\
      & \\
      &  p_{{k},j}\geq \beta\cdot \mu(\ell_{{k},j})v_{{k},j} \quad \forall j\in[M]\,,
\end{align*}
where the first inequality follows from bidder ${k}$'s ROAS constraint; the second  inequality follows from the fact that any winning bidder's payment must be greater than her $\beta$-approximate reserves. Combining the above inequalities and rearranging we get
\begin{align}
\label{eq:VCG:badbidder2}
   \sum_{j\in \mathcal{B}_{i}(k;\bm{b})} p_{{k},j} \leq \sum_{j\in \mathcal{B}_{i}(k;\bm{b})}\mu(\ell_{{k},j})v_{{k},j} + (1-\beta)\cdot \sum_{j\notin \mathcal{B}_{i}(k;\bm{b})}  \mu(\ell_{{k},j})v_{{k},j}\,,
\end{align}
Summing Eq.\eqref{eq:VCG:badbidder1} over all $j\in \mathcal{B}_{i}(k;\bm{b})$ and combining with Eq. \eqref{eq:VCG:badbidder2}, we get
\begin{align*}
\begin{aligned}
  & \left(\alpha_{i}-1\right)\cdot \sum_{j\in \mathcal{B}_{i}(k;\bm{b})}\left(\mu(\ell_{i,j}^{*}) - \mu(\ell_{i,j})\right) v_{i,j}
 \\
 ~\leq~ &  (1- \beta) \cdot \left( \sum_{j\in \mathcal{B}_{i}(k;\bm{b})}\mu(\ell_{i,j})v_{i,j} +  \sum_{j\notin \mathcal{B}_{i}(k;\bm{b})}  \mu(\ell_{{k},j})v_{{k},j}\right) + \sum_{j\in \mathcal{B}_{i}(k;\bm{b})}\mu(\ell_{{k},j})\left(v_{{k},j} -v_{i,j}\right)\\
  ~\overset{(a)}{\leq}~ & (1- \beta) \cdot \left( \sum_{j\in \mathcal{B}_{i}(k;\bm{b})}\mu(\ell_{i,j})v_{i,j} +  \sum_{j\notin \mathcal{B}_{i}(k;\bm{b})}  \mu(\ell_{{k},j})v_{{k},j} + \sum_{j\in \mathcal{B}_{i}(k;\bm{b})}\mu(\ell_{{k},j})\left(v_{{k},j} -v_{i,j}\right)\right)\\
  ~\overset{(b)}{\leq}~ & (1- \beta) \cdot \left( \sum_{j\in \mathcal{B}_{i}(k;\bm{b})}\mu(\ell_{i,j})v_{i,j} +  \sum_{j\in[M]}  \mu(\ell_{{k},j})v_{{k},j}  - \sum_{j\in \mathcal{B}_{i}(k;\bm{b})}\mu(\ell_{i,j}^{*})v_{i,j}\right)\,.
  \end{aligned}
\end{align*}
In (a), we used the fact that $\beta\in (0,1]$ and $v_{{k},j} -v_{i,j}< 0$ for any $k\in \mathcal{C}\subseteq \mathcal{C}_{i}(\bm{\outcome})$; see definition of $\mathcal{C}_{i}(\bm{\outcome})$ in Eq. \eqref{eq:covering}; and (b) follows from $\ell_{{k},j} \leq \ell_{i,j}^{*}$ for any $k\in \mathcal{C}\subseteq \mathcal{C}_{i}(\bm{\outcome})$. Rearranging terms we obtain the desired Eq. \eqref{eq:welfarelossdue2k}. \halmos

\subsubsection{A tighter individual welfare guarantee}

We recognize that the individual welfare lower bound guarantee in Theorem \ref{thm:fairnessbound}, namely $1- \frac{1-\beta}{\alpha_{i}-1} \cdot \frac{\opt_{-i}}{\opt_{i}}$, may be negative and hence meaningless for a small advertiser, i.e. advertiser $i$ whose market share $\frac{\opt_{i}}{\opt_{-i}}$ is very small which may be a result of a very large number of bidders $N$. Nevertheless, in light of the proof above for Theorem \ref{thm:fairnessbound}, we can in fact present a tighter individual welfare guarantee that replaces $\opt_{-i}$ in the numerator, i.e. total welfare summed over $N-1$ competitors of bidder $i$, by the total welfare of a potentially much smaller subset of bidder $i$'s competitors. 

Analogous to the definitions $\mathcal{L}_{i}(\bm{\outcome}), \mathcal{B}_{i}(k; \bm{\outcome})$ and $\mathcal{C}_{i}(\bm{\outcome})$ for any outcome $\bm{\outcome}$ defined in Definition \ref{def:welfareloss}, Eqs. \eqref{eq:covering0} and \eqref{eq:covering}, respectively, we slightly abuse notation and define their outcome-independent counterparts.
\begin{align}
\begin{aligned}
\label{eq:gencovering}
& \mathcal{L}_{i} = \left\{j\in [M]: \opt_{i,j} > 0\right\}\\
 & \mathcal{B}_{i}(k)= \left\{j\in [M]: \opt_{i,j} > 0,~ v_{k,j} < v_{i,j}\right\}\\
   & \mathcal{C}_{i} = \left\{\mathcal{C} \subseteq [N]/\{i\}: (\mathcal{B}_{i}(k))_{k\in \mathcal{C}} \text{ is a maximal set cover of }\mathcal{L}_{i}\right\}\,,
\end{aligned}
\end{align}
where we recall $\opt_{i,j}$ is the welfare of bidder $i$ in auction $\auc_{j}$ under the efficient outcome as defined in Eq. \eqref{eq:optefficient}. 
In words, $\mathcal{L}_{i}$ is the set of auctions in which bidder $i$ can potentially lose welfare, $\mathcal{B}_{i}(k)$ is the set of auctions in which competitor $k$ can potentially cause $i$ to lose welfare, and any covering of competitors $\mathcal{C} \in  \mathcal{C}_{i}$ can potentially cause $i$ to lose welfare in all auctions of $ \mathcal{L}_{i} $. 

We remark that $ \mathcal{C}_{i} $ only depends on bidder values $(v_{i,j})_{i\in [N],j \in [M]}$, and it is easy to see that any covering $\widetilde{\mathcal{C}} \in \mathcal{C}_{i}$ has cardinality at most $\min\{ M, N-1\}$.  To exemplify the covering set $ \mathcal{C}_{i}$, consider an instance consisting of 2 single slot VCG auctions (i.e. SPA)  and 3 bidders with the following advertiser values
\vspace{0.5cm}
\renewcommand{\arraystretch}{1.5}
\begin{center}
\begin{tabular}{|c | c c c| } 
 \hline
  & SPA 1 & SPA 2 & SPA 3   \\
 \hline\hline
 bidder 1 & $v_{1,1} = 2$ & $v_{1,2} = 5$ & $v_{1,3} = 0$ \\
 \hline
 bidder 2 & $v_{2,1} = 1$  & $v_{2,2} =1$ & $v_{2,3} =10$ \\
 \hline
  bidder 3 & $v_{3,1} = 0$  & $v_{3,2} =4$ & $v_{3,3} = 10$\\
 \hline
\end{tabular}
\end{center}
\medskip
Then under the efficient outcome, bidder 1 wins auctions 1 and 2. However, it may be possible that bidder 2 solely outbids the other bidders to win both auctions 1 and 2, inducing a covering $\{2\}$ for bidder 1; or bidder 2 wins auction 1 while bidder 3 wins auction 2, inducing a covering $ \{2,3\}$  for bidder 1. Therefore, the covering set $\mathcal{C}_{i} = \{ \{2\}, \{2,3\}\}$.

The following proposition states that any covering $\mathcal{C} \in \mathcal{C}_{i}(\bm{\outcome})$ that contributes to all bidder $i$'s welfare loss under $\bm{\outcome}$, must be a subset of an element of the set $\mathcal{C}_{i}$.
\begin{proposition}
\label{prop:coveringsubset}
    Let $\bm{\outcome} $ be any outcome and denote $\mathcal{C}_{i}(\bm{\outcome})$ be the corresponding set of coverings defined in Eq. \eqref{eq:covering}. Then, for any 
    covering $\mathcal{C}\in \mathcal{C}_{i}(\bm{\outcome})$, there must exist an $\widetilde{\mathcal{C}} \in  \mathcal{C}_{i}$ such that $\mathcal{C}\subseteq \widetilde{\mathcal{C}}$, where $\mathcal{C}_{i}$ is defined in Eq. \eqref{eq:gencovering}.
\end{proposition}

In the following theorem, we present a tighter individual welfare guarantee than that of Theorem \ref{thm:fairnessbound} by replacing $\opt_{-i}$ with the welfare of some covering for bidder $i$, which may potentially be a very small subset of all bidder $i$'s competitors that includes at most $\min\{M,N-1\}$ bidders.
\begin{theorem}
\label{thm:improvedbound}
    Consider $(\auc_{j})_{j\in [M]}$ are VCG auctions, and personalized reserve prices $\bm{r}$ are $\beta$-approximate as in Definition \ref{def:reserve}.
Fix an autobidder $i \in [K]$ who adopts bid multiplier $\alpha_{i} > 1$ (see Proposition \ref{prop:ubid}) so $\bm{b}_{i} = \alpha_{i} \bm{v}_{i}$. Then, the individual welfare guarantee in Definition \ref{def:fairness}) is bounded as:
\begin{align*}
  \min_{\bm{b}_{-i}\in \feas_{-i}(\alpha_{i}\bm{v}_{i})}  \frac{\well_{i}(\alloc(\bm{b}))}{\opt_{i}} ~\geq~ 1- \frac{1-\beta}{\alpha_{i}-\beta} \cdot \frac{\max_{\widetilde{\mathcal{C}}\in \mathcal{C}_{i}} \sum_{k \in \widetilde{\mathcal{C}}}\well_{k}(\alloc(\bm{v}_{\widetilde{\mathcal{C}}},\bm{0}))}{\opt_{i}}\,,
\end{align*}
where $\feas_{-i}(\cdot)$ is defined in Definition \ref{def:feas}, $\bm{v}_{\widetilde{\mathcal{C}}} = (\bm{v}_{k})_{k\in \widetilde{\mathcal{C}}}$, and $\alloc(\bm{v}_{\widetilde{\mathcal{C}}},\bm{0})$ is the outcome when bidders in covering $\widetilde{\mathcal{C}}$ bid truthfully while others bid 0; equivalently, this is the total efficient welfare when participation in all auctions are restricted to bidders in $\widetilde{\mathcal{C}}$ only.
\end{theorem}

\textit{Proof.}
Let $\bm{b}\in \feas$ be any feasible bid profile, and let $\bm{\outcome} = \alloc(\bm{b})$ be the corresponding outcome. Also, let $\mathcal{C}_{i}(\bm{b})$ be the set of coverings defined in Eq. \eqref{eq:covering}, and consider any  $\mathcal{C} \in \mathcal{C}_{i}(\bm{\outcome})$. In Eq. \eqref{eq:vcgboundloss} within the proof of Theorem \ref{thm:fairnessbound}, we showed $\loss_{i}(\bm{\outcome}) ~\leq~
  \frac{1-\beta}{\alpha_{i}-1}\sum_{k \in \mathcal{C}}\well_{k}(\bm{\outcome})$, so
\begin{align*}
  \loss_{i}(\bm{\outcome}) ~\leq~&
  \frac{1-\beta}{\alpha_{i}-1}\sum_{k \in \mathcal{C}}\well_{k}(\bm{\outcome}) 
  ~\overset{(a)}{\leq}~
  \frac{1-\beta}{\alpha_{i}-1}\sum_{k \in \widetilde{\mathcal{C}}}\well_{k}(\bm{\outcome})\\
   ~\leq~&
  \frac{1-\beta}{\alpha_{i}-1}\max_{\widetilde{\mathcal{C}}\in \mathcal{C}_{i}} \sum_{k \in \widetilde{\mathcal{C}}}\well_{k}(\alloc(\bm{v}_{\widetilde{\mathcal{C}}},\bm{0}))
\,,
\end{align*}
where in (a) we let $\widetilde{\mathcal{C}} \in \mathcal{C}_{i}$ defined in Eq. \eqref{eq:gencovering} such that $\mathcal{C} \subseteq \widetilde{\mathcal{C}}$ according Proposition \ref{prop:coveringsubset}.
Rearranging and applying Proposition \ref{prop:boundloss} yields the desired lower bound.
\halmos

We conclude by making the remark that although the individual welfare lower bound in Theorem \ref{thm:improvedbound}  may potentially be stronger that in Theorem \ref{thm:fairnessbound}, it comes at the cost of significantly increased computational complexity due to the maximum over all coverings. Nevertheless, this improved bound may still be practical for instances with relatively small number of auctions, since the cardinality of any covering is upper bounder by the number of auctions. 

\subsection{Applicability of the individual welfare guarantee when all bidders bid uniformly}
We recognize that as the individual welfare lower bound in Theorem \ref{thm:fairnessbound} monotonically increases in the  bid multiplier $\alpha_{i}$, it is tempting for bidder $i$ to apply a very large multiplier $\alpha_{i}$. Nevertheless, in this section we describe a potential tradeoff between large multipliers (i.e. better individual welfare guarantees in light of Theorem \ref{thm:fairnessbound}) and ROAS feasibility in the practical scenario where all bidders are autobidders and adopt uniform bidding. 

To illustrate, we see that for large multiplier $\alpha_{i}$, the set of competing bids $\feas_{-i}(\alpha_{i}\bm{v}_{i})$ may only include very small bid values (e.g. the bid profile where each competing bidder (under)bids some small $\epsilon > 0$ close to 0 in each auction), at which bidder $i$ faces nearly no competition so that the ROAS constraint can be trivially satisfied for every bidder. In light of this discussion,  we  consider a more practical scenario where all competing bidders are also autobidders and adopt uniform bidding, or equivalently, a refinement of $\feas_{-i}(\alpha_{i}\bm{v}_{i})$ 
 in which each competing bidder $j\neq i$, similar to bidder $i$, also adopts uniform bidding with bid multiplier $\alpha_{j}\geq 1$. We define
 $\feas_{-i}^{u}(\bm{b}_{i}) = \feas_{-i}(\bm{b}_{i})\cap \{(\alpha_{j}\bm{v}_{j})_{j\neq i}: \alpha_{j}\geq 1\}$ that represents
the set of  uniform competing bids for bidder $i$ that ensure ROAS constraint satisfaction for every bidder. From Theorem \ref{thm:fairnessbound}, it is easy to see
\begin{align}
\begin{aligned}
\label{eq:ubidfairness}
   & \min_{\bm{b}_{-i}\in \feas_{-i}^{u}(\alpha_{i}\bm{v}_{i})}  \frac{\well_{i}(\alloc(\bm{b}))}{\opt_{i}} ~\overset{(i)}{\geq} ~ 1- \frac{1-\beta}{\alpha_{i}-1} \cdot \frac{\opt_{-i}}{\opt_{i}}\,,
\end{aligned}
\end{align}
where (i) follows from $\min_{\bm{b}_{-i}\in \feas_{-i}^{u}(\alpha_{i}\bm{v}_{i})}  \frac{\well_{i}(\alloc(\bm{b}))}{\opt_{i}} \geq \min_{\bm{b}_{-i}\in \feas_{-i}(\alpha_{i}\bm{v}_{i})}  \frac{\well_{i}(\alloc(\bm{b}))}{\opt_{i}} $ because $\feas_{-i}^{u}(\alpha_{i}\bm{v}_{i}) \subseteq \feas_{-i}(\alpha_{i}\bm{v}_{i})$.
Nevertheless,  in light of Eq. \eqref{eq:ubidfairness}, when all bidders bid uniformly, an excessively large $\alpha_{i}$ may let bidder $i$ incur  large payments that significantly exceed her values, resulting in non-existence of competing uniform bids $\bm{b}_{-i}$ that can ensure satisfaction of every bidders' ROAS constraints, i.e. $\feas_{-i}^{u}(\alpha_{i}\bm{v}_{i})$ being empty.  In other words, there exists a tradeoff between large multipliers (i.e. better individual welfare guarantees) and ROAS feasibility when all bidders bid uniformly. The following Lemma \ref{lem:validmultiregion}, along with a technical definition of ``well-separated'' values per Definition \ref{def:separate}, addresses this tradeoff by characterizing how large the multiplier $\alpha_{i}$ can be set that still ensures the existence of uniform competing bids within $\feas_{-i}^{u}(\alpha_{i}\bm{v}_{i})$.
\begin{definition}[$\Delta$-separated values]
\label{def:separate}
We say values $\bm{v}\in \R_{\geq 0}^{N\times M}$ are $\Delta$-separated for some $\Delta > 1$ if  any value $v_{i,j}$ is at least $\Delta$ times as much as any value that is less than $v_{i,j}$ in the same auction $j$, i.e. $v_{i,j}\geq \Delta\cdot \max\{v_{k,j}: k\in [N], v_{k,j}< v_{i,j}\}$ for any bidder $i$ and auction $j$.\footnote{
 Definition \ref{def:separate} also captures values which are ``additively separated''. In particular, take some $d >0$ such that $d < \min\{v_{i,j}: v_{i,j}\neq 0\}$ and also
$v_{i,j} -d \geq \max\{v_{k,j}: k\in [N], v_{k,j}< v_{i,j}\}$ for any bidder $i$ and auction $j$. Then, by taking $\Delta \in \min_{v_{i,j}:v_{i,j}\neq 0}\left\{\frac{v_{i,j}}{v_{i,j}-d}\right\}$, the values are $\Delta$-separated according to Definition \ref{def:separate} because $\frac{1}{\Delta}v_{i,j} \geq v_{i,j} -d \geq \max\{v_{k,j}: k\in [N], v_{k,j}< v_{i,j}\}$ for all $v_{i,j}$. 
This suggests Definition \ref{def:separate}  is quite general to capture value separation scenarios.}
\end{definition}

\begin{figure*}[t!]
    \centering
    \begin{subfigure}[t]{0.45\linewidth}
        \centering
        \includegraphics[width=\linewidth]{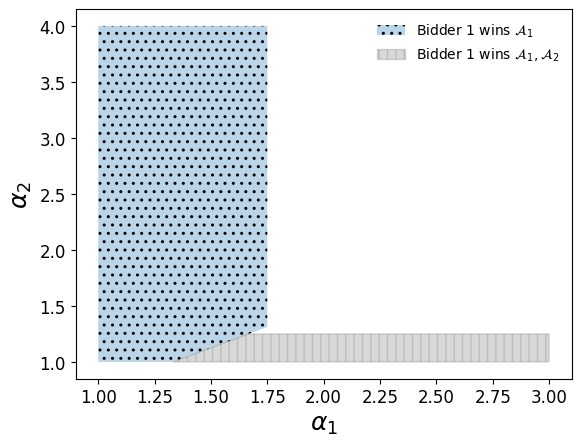}
    \end{subfigure}%
    ~ 
    \begin{subfigure}[t]{0.45\linewidth}
        \centering
        \includegraphics[width=\linewidth]{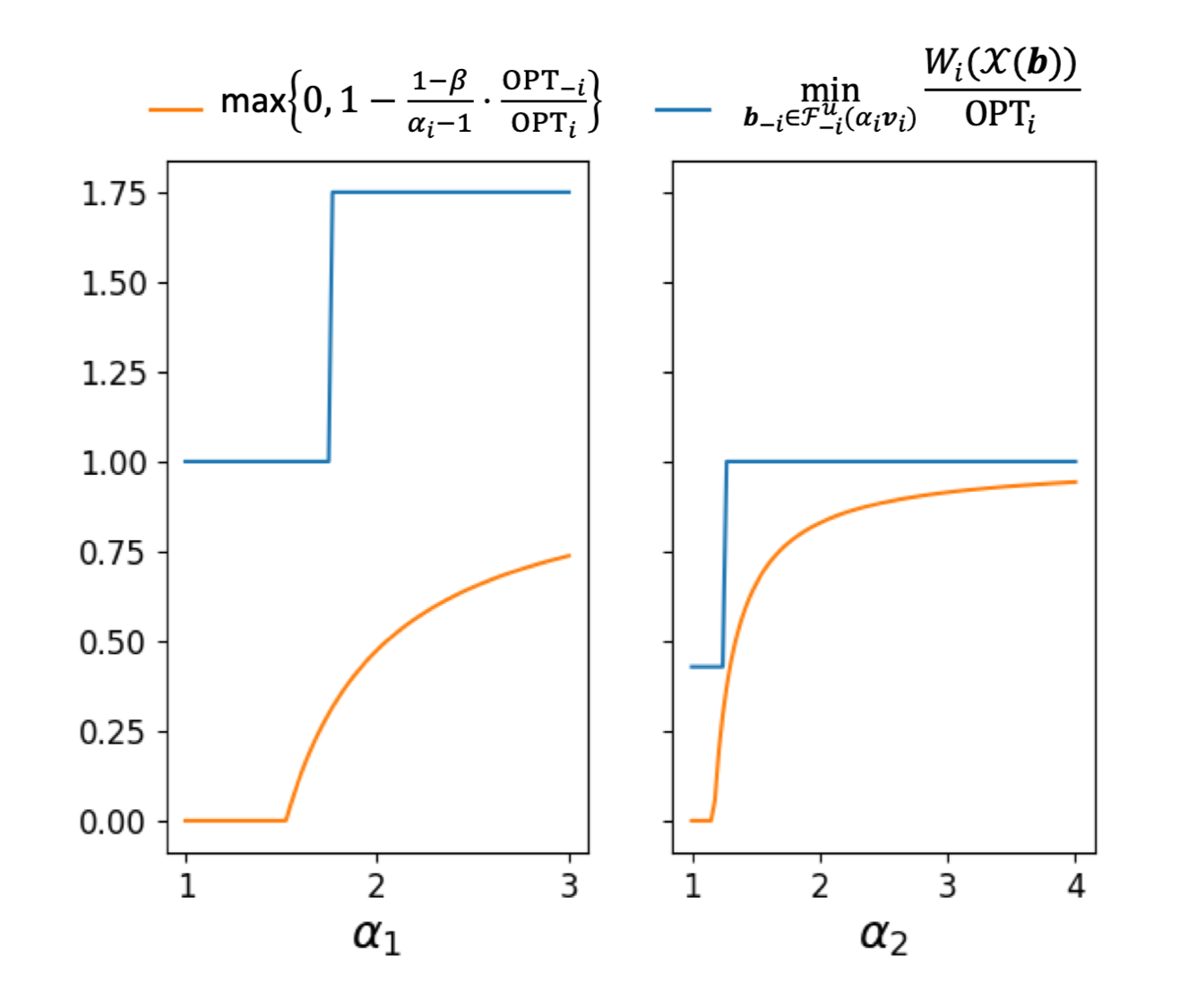}
    \end{subfigure}
    \caption{
    Left: Two colored regions represent uniform bid multipliers $(\alpha_{1},\alpha_{2})\in [1,\infty)^{2}$ that lead to feasible bid profiles $(\bm{b}_{1},\bm{b}_{2}) \in \feas$. Right: Comparison between the individual welfare guarantee of Theorem \ref{thm:fairnessbound}, namely $1-\frac{1-\beta}{\alpha_{i}-1}\cdot \frac{\opt_{-i}}{\opt_{i}}$, and the worst case welfare for each bidder $i$ (normalized by $\opt_{i}$) among all feasible bid profiles when both bidders adopt uniform bidding, namely  $\min_{\bm{b}_{-i}\in \feas_{-i}^{u}(\alpha_{i}\bm{v}_{i})}  \frac{\well_{i}(\alloc(\bm{b}))}{\opt_{i}} $.}
    \label{fig:compare}
\end{figure*}

\begin{lemma}[Valid regions for uniform bid multiplier]
\label{lem:validmultiregion}
Let $(\auc_{j})_{j\in[M]}$ be VCG auctions and assume bidders values are $\Delta$-separated (Definition \ref{def:separate}) in every auction for some $\Delta > 1$, then $\feas_{-i}^{u}(\alpha_{i}\bm{v}_{i}) \neq \varnothing$ for any $\alpha_{i}\in \left[1, \Delta \right)$.
\end{lemma}
The proof of this lemma is presented in Appendix \ref{pf:lem:validmultiregion}. We also remark that the upper bound $\Delta$ in Lemma \ref{lem:validmultiregion} is sufficient, meaning that there may exist larger values of $\alpha_{i}$ that can ensure the set $\feas_{-i}^{u}(\alpha_{i}\bm{v}_{i}) \neq \varnothing$ nonempty. To better visualize the structure of $\feas_{-i}^{u}(\alpha_{i}\bm{v}_{i})$, as well as our individual welfare guarantee in Theorem \ref{thm:fairnessbound} and Eq. \eqref{eq:ubidfairness}, we present the following example.
\begin{example}
    Consider 2 bidders bidding in 3 single-slot VCG auctions in which each slot is associated with CTR equal to 1. Bidder values are $\bm{v}_{1}= (4,3,1)$ and $\bm{v}_{2}= (1,4,3)$, while personalized reserves are set to be $\bm{r}_{i} = \beta \bm{v}_{i}$ for $\beta = 0.7$ and $i = 1,2$. It is easy to check that with the presence of personalized reserves, no bidder can significantly overbid and win all auctions (otherwise she will incur large payments and thus violate their ROAS constraints), and therefore each bidder will obtain non-zero value. This aligns with our intuition presented in Sections \ref{sec:reserves} that states personalized reserves benefit individual welfare. 
    
    In the left subgraph of Figure \ref{fig:compare}, we color the region of all pairs of uniform bid multipliers $(\alpha_{1},\alpha_{2})\in [1,\infty)^{2}$ that induce feasible bid profiles $(\bm{b}_{1},\bm{b}_{2}) \in \feas$, where the blue dotted region corresponds to bid profiles under which bidder 1 wins only $\auc_{1}$, and the grey vertically-dashed region corresponds to bid profiles under which bidder 1 wins $\auc_{1}$ and $\auc_{2}$. From this subgraph, we can see that $\feas_{-1}^{u}(\alpha_{1}\bm{v}_{1})= \{\alpha_{2}\bm{v}_{2}: \alpha_{2}\in \text{any colored vertical line segments at } \alpha_{1}\}$ and similarly $\feas_{-2}^{u} (\alpha_{2}\bm{v}_{2})= \{\alpha_{1}\bm{v}_{1}: \alpha_{1}\in \text{any colored horizontal line segments at } \alpha_{2}\}$. On the right subgraph of Figure \ref{fig:compare}, for each bidder $i = 1,2$, we plot the individual welfare guarantee $1-\frac{1-\beta}{\alpha_{i}}\cdot \frac{\opt_{-i}}{\opt_{i}}$ as well as 
$\min_{\bm{b}_{-i}\in \feas_{-i}^{u}(\alpha_{i}\bm{v}_{i})}  \frac{\well_{i}(\alloc(\bm{b}))}{\opt_{i}} $ which is the worst case welfare among all outcomes induced by uniform bid profiles that satisfy both bidders' ROAS constraints. 
\end{example}
On the left subgraph of Figure \ref{fig:compare}, we observe that it is easier for bidder 2 to ensure a non-empty feasibility set $\feas_{-2}^{u} (\alpha_{2}\bm{v}_{2})$ at large $\alpha_{2}$ values than bidder 1 to ensure non-empty $\feas_{-1}^{u} (\alpha_{1}\bm{v}_{1})$ at large $\alpha_{1}$; e.g. for large $\alpha_{2}$ such as $\alpha_{2} = 3$, bidder 1 can take any $\alpha_{1} \in [1,1.75]$, but for large $\alpha_{1} = 3$, bidder 2 can only take $\alpha_{2} \in [1,1.25]$. Nevertheless on the right subgraph, we see that bidder 2's realized welfare is much closer to her theoretical lower bound guarantee than that of bidder 1. Therefore this highlights a tradeoff between uniform multiplier feasibility and welfare guarantee.

\section{Impossibility result: VCG yields best individual welfare guarantee}
\label{sec:imposs}
Having presented an individual welfare guarantee in the previous Section \ref{sec:VCG} that improves according to the platform's ML advice accuracy, a natural question is that for a given level of accuracy $\beta$, can one achieve a universally better individual welfare guarantee than that of Theorem \ref{thm:fairnessbound} via considering auction formats other than VCG? In this section, we demonstrate that the answer is negative when we restrict the auction to a broad class of truthful mechanisms (possibly randomized) with anonymous allocations (see Definition \ref{def:anonalloc}). Here, we again emphasize that truthfulness is w.r.t. quasi-linear utility maximizers (see Definition \ref{def:truthfulauctions}).

In the following theorem, we  show that no allocation-anonymous, truthful auction $\auc$
when augmented by $\beta$-approximate reserves (see Definition \ref{def:resaugment}),  can universally outperform VCG, i.e. for any $\auc$ there exists a problem instance in which a bidder has a welfare guarantee at most the individual welfare lower bound for VCG of Theorem \ref{thm:fairnessbound}.
\begin{theorem}
\label{thm:imposs}
Let $\auc$ be any single-slot auction format (with position bias $\mu = 1$) that is allocation-anonymous, truthful, and possibly randomized. Then, there exists an instance of $M$ parallel auctions $(\auc_{j})_{j\in[M]}$ of format $\auc$,  $N$ bidders with values $\bm{v}\in \R_{+}^{N\times M}$, $\beta$-approximate reserves $\bm{r}\in \R_{+}^{N\times M}$, and an autobidder $i$ with multiplier $\alpha_{i} > 1$ (see Proposition \ref{prop:ubid}), such that there is a feasible bid profile $\bm{b}\in \feas$ in which $\bm{b}_{i} = \alpha_{i}\bm{v}_{i}\in \R_{+}^{M}$ that results in the following welfare upper bound for autobidder $i$: $\frac{\E\left[\well_{i}(\alloc(\bm{b}))\right]}{\E\left[\opt_{i}\right]}\leq 1 - \frac{1-\beta}{\alpha_{i}-1}\cdot \frac{\E\left[\opt_{-i}\right]}{\E\left[\opt_{i}\right]}$ 
 . Here the expectation is taken w.r.t. possible randomness in $\mathcal{A}$.
\end{theorem}
Details on implementation allocation-anonymous auctions with personalized reserve prices can be found in Definition \ref{def:resaugment} and Example \ref{ex:augmentstandard}. Our proof strategy for Theorem \ref{thm:imposs} is to construct a ``bad'' autobidding instance for any auction $\mathcal{A}$ of interest that yields low individual welfare for one specific bidder: we show that in this autobidding instance, there is some bidder $i$ who has a welfare upper bound as stated in the theorem. The construction of this bad autobidding instance is motivated by Example \ref{ex:motivate}, in which the key source of low welfare for an individual bidder $i$ comes from the fact that competing bidders outbid $i$ in auctions where $i$'s value is high, and cover their expenditures with value acquired from other auctions where they have no competition. Following this idea, since the bad instance in Theorem \ref{thm:imposs} requires us to maximize individual welfare for a specific bidder $i$, we can achieve this by having auctions where each of $i$'s competitors is the only bidder submitting a nonzero bid, and with these ``no-competition'' auctions competitors can cover their expenditures for outbidding bidder $i$ in auctions where $i$'s value is largest.

\textbf{Proof sketch for Theorem \ref{thm:imposs}. } For any auction $\mathcal{A}$ that is allocation-anonymous, truthful and possibly randomized, we consider a ``bad'' autobidding instance $(N,M,\mathcal{A},\bm{r},\bm{v})$ where $N = K+1$ bidders labeled $B_{1}...B_{K},B_{0}$ compete in $M = 2K+1$ auctions with single-slots for some $K\in\N$, and bidders' values are shown in the following table. Reserves are set to be $r_{i,j}= \beta v_{i,j}$
for some $\beta \in (0,1)$ and are $\beta$-approximate (see Definition \ref{def:reserve}). 
Bidder $B_{0}$'s multiplier is fixed at $\alpha_{0} > 1$.
\vspace{0.5cm}
\renewcommand{\arraystretch}{1.5}
\begin{center}
\begin{tabular}{|c | c c c c c c c c c|} 
 \hline
  & $A_{1}$ & $A_{2}$ & \dots  &   $A_{K}$  & $A_{K+1}$ & $A_{K+2}$ & \dots & $A_{2K}$ &  $A_{2K+1}$\\
 \hline\hline
$B_{1}$  & $\frac{\alpha_{0}v+\epsilon}{\rho}$ & $\frac{\alpha_{0}v+2\epsilon}{\rho}$  & \dots &  $\frac{\alpha_{0}v+K\epsilon}{\rho}$ & $\gamma$  & 0 &  \dots & 0 & 0\\
  \hline
$B_{2}$   & $\frac{\alpha_{0}v+2\epsilon}{\rho}$ & $\frac{\alpha_{0}v+3\epsilon}{\rho}$  & \dots &  $\frac{\alpha_{0}v+\epsilon}{\rho}$ & 0 &  $\gamma$ & \dots &  0 & 0  \\
 \hline
 \vdots & \vdots &  \vdots  &  &   \vdots  &  \vdots & \vdots & & \vdots & \vdots \\
  \hline
$B_{K}$  & $\frac{\alpha_{0}v+K\epsilon}{\rho}$ & $\frac{\alpha_{0}v+\epsilon}{\rho}$  & \dots &  $\frac{\alpha_{0}v+(K-1)\epsilon}{\rho}$ & 0  & 0  & \dots & $\gamma$ & 0\\
   \hline
$B_{0}$  & $v$ & $v$ & \dots & $v$ & 0 & 0 & \dots & 0  & ${y}$\\ 
 \hline
\end{tabular}
\end{center}
\vspace{0.5cm}
In the table, we choose $\epsilon = \mathcal{O}(1/K^{3})$ and suitable parameters $\rho, \gamma,v,{y} > 0$ to satisfy certain conditions, one of which guarantees $B_{0}$'s value is the highest in auctions $A_{1} \ldots A_{K}$ . With the above instance, we consider the specific outcome $\bm{x}$ where bidders $1,\ldots, K$ adopt bid multiplier $\rho$, in which case bidder $B_{0}$ has the lowest bid in auctions $A_{1} \ldots A_{K}$ (since $\alpha_{0} > 1$. Then, our proof of Theorem \ref{thm:imposs} 
is to show bidder $B_{0}$ can acquire welfare  at most the upper bound in Theorem \ref{thm:imposs}. The proof consists of 3 parts: 

(1) Under outcome $\bm{x}$, we upper bound bidder $B_{0}$'s expected acquired welfare in auctions $A_{1} ... A_{K}$. This acquired welfare should be small, since other bidders are outbidding bidder $B_{0}$ in these auctions, by covering their expenditures via the value acquired in auctions $A_{K+1},...,A_{2K}$, respectively. 

(2) We show that bidder $B_{0}$ satisfies her ROAS constraint, which holds valid due to the fact that she is acquiring value in auction $A_{2K+1}$ for suitable ${y}$ facing no competition.

(3) We show that any bidder $i\in [K]$ satisfies her ROAS constraint. In this part, when we take 
appropriate parameters $\epsilon\to 0$ and
$\rho \to \alpha_{0}$, we first show that the total expected acquired value of bidder $i$ minus her total expected cost over all $2K+1$
 auctions is approximately $1-\sum_{j\in [K]}\prob\left(\text{bidder $i$ wins auction $j$}\right) + \mathcal{O}(K^{2}\epsilon)$ and since $\epsilon = \mathcal{O}(1/K^{3})$, we only need to show $1\geq\sum_{j\in [K]}\prob\left(\text{bidder $i$ wins auction $j$}\right)$.  Our proof for this claim exploits  the specific structure of our problem instance construction: we recognize that when bidders $1,\ldots,K$ use bid multiplier $\rho$, the bid profiles for auctions $A_{1} ... A_{K}$ are a cyclic permutation of the set $\{b_{0},\ldots,b_{K}\} = \{\alpha_{0}v,\alpha_{0}v + \epsilon, \ldots, \alpha_{0}v+K\epsilon\}$. Therefore by allocation-anonymity of $\mathcal{A}$, the expected outcome in each of the auctions in $A_{1}, \ldots, A_{K}$ are symmetric over bidders $1, \ldots, K$, or in other words,
$\prob\left(\text{bidder $i$ wins auction $j$}\right)$ depends only on the bid value of bidder $i$ in auction $j$. The cyclic structure thus implies 
\begin{align*}
\begin{aligned}
    & \sum_{j\in [K]}\prob\left(\text{bidder $i$ wins auction $j$}\right) \\
    = &  \sum_{k\in [K]}\prob\left(\text{bid value $b_{k}$ wins auction $\mathcal{A}$ given competing bids $b_{-k}$}\right) \leq 1\,.
\end{aligned}
\end{align*}
Here, we also point out that any $\beta$ approximate reserves do not affect allocation in  auctions $A_{1}...A_{K}$, simply because any bid value in $\{b_{0},\ldots,b_{K}\}$ is greater than the largest reserve price among agents, namely $\beta v$, 
since $\alpha_{0}> 1 > \beta$. In other words, under the specific outcome $\bm{x}$, allocation anonymity of any auction in $A_{1},\ldots,A_{K}$ is preserved with personalized reserves $r_{i,j}=\beta v_{i,j}$ due to our construction; see also Remark \ref{rmk:approxresanon}. For a technical re-statement of Theorem \ref{thm:imposs} and its proof, please refer to Appendix \ref{pf:thm:imposs}.

\section{Extensions: Individual welfare guarantees for GSP and GFP with ML advice}
\label{sec:extensions}

In this section, we extend our individual welfare guarantees for the VCG auction in Theorem \ref{thm:fairnessbound} to the GSP and GFP auctions, which are both non-truthful.  For technical purposes, we assume that bidder values are ``well-separated'' as defined in Definition \ref{def:separate}.


Further, as discussed in Section \ref{subsec:autobidder}, uniform bidding (i.e. setting the same bid multiplier for all auctions) is only optimal in truthful auctions. In GSP and GFP, one can construct instances where non-uniform bidding strictly outperforms uniform bidder (for more details see e.g. \cite{deng2019prior}). Thus, for GSP and GFP autobidding instances, we impose no assumptions on the bid values of bidders other than being undominated: we say a bid value $\bm{b}_{i}\in \R_{+}^{M}$ is undominated for bidder $i$ if there is no other bid value $\bm{b}_{i}'\in \R_{+}^{M}$ that strictly outperforms  $\bm{b}_{i}$ in welfare under all competing bid profiles. Mathematically, $\nexists \bm{b}_{i}'\in \R_{+}^{M}$ such that $\well_{i}(\alloc(\bm{b}_{i},\bm{b}_{-i})) < \well_{i}(\alloc(\bm{b}_{i}',\bm{b}_{-i}))$ for all $\bm{b}_{-i}\in \R_{+}^{(N-1)\times (M)}$. The following lemma lower bounds undominated bids in the presence of $\beta$-approximate reserves.
\begin{lemma}[Lemma 4.7 \& 4.9 of \cite{balseiro2021robust}]
\label{lem:undominatedbids}
Consider the setting where  $(\auc_{j})_{j\in[M]}$  are all  GSP auctions or GFP auctions, and reserve prices $\bm{r}$ are $\beta$-approximate. Denote $\mathcal{U} \subseteq \feas$ to be the set of bid profiles in which each bid is undominated and satisfies all bidders' ROAS constraints. Then for any $\bm{b}\in \mathcal{U}$, $b_{i,j}$ must satisfy $b_{i,j}\geq r_{i,j}\geq \beta v_{i,j}$ for any bidder $i \in [N]$ and auction $\auc_{j}$. 
\end{lemma}

Finally, our main theorem for this section is the following:
\begin{theorem}
\label{thm:GFP:fairnessbound}
Consider the setting where  $(\auc_{j})_{j\in[M]}$ are all GSP auctions or GFP auctions. Suppose reserve prices $\bm{r}$ are $\beta$-approximate, and values $\bm{v}$ are $\Delta$-separated s.t. $\beta>\frac{\Delta}{2\Delta-1}$. Consider any undominated bid profile $\bm{b}\in \mathcal{U} \subseteq \feas$ where $\mathcal{U} $ is the set of all undominated bids under which every bidder's ROAS constraint is satisfied (see Equation (\ref{eq:autobidderprob})). Then,  the individual welfare guarantee (Definition \ref{def:fairness}) is bounded as
\begin{align*}
   \min_{\bm{b}\in \mathcal{U}}  \frac{\well_{i}(\alloc(\bm{b}))}{\opt_{i}} ~\geq~ 1- \frac{1-\beta}{\beta -\frac{\Delta}{2\Delta-1}} \cdot \frac{\opt_{-i}}{\opt_{i}}\,.
\end{align*}
\end{theorem}
\vspace{-0.1cm}
Details on implementation of GSP and GFP with personalized reserve prices can be found in Definition \ref{def:resaugment} and Example \ref{ex:augmentstandard}. The proof for Theorem \ref{thm:GFP:fairnessbound} is presented in Appendix \ref{pf:thm:GFP:fairnessbound}. Comparing the individual welfare guarantees in Theorem \ref{thm:fairnessbound} for VCG and Theorem \ref{thm:GFP:fairnessbound} for GSP/GFP, we observe when values are $\Delta$-separated and ML advice is $\beta$-accurate, when bidders adopt small enough uniform multipliers in VCG (i.e. $\alpha_{i} -1< \beta -\frac{\Delta}{2\Delta-1} $),  GSP/GFP provides a better individual welfare guarantee compared to VCG, whereas for large multipliers (i.e. $\alpha_{i}-1 > \beta - \frac{\Delta}{2\Delta-1}$), individual welfare in VCG dominates that in the considered non-truthful auctions.

\section{Numerical studies}
\label{sec:numerics}
Recall in Sections \ref{sec:VCG} and \ref{sec:extensions} we theoretically showed that setting personalized reserve prices with ML advice (see Definition \ref{def:reserve}) ensures worst-case welfare guarantees on the individual bidder level in VCG, GSP, GFP auctions, and moreover, such guarantees improves as ML advice becomes more accurate. In this section, we complement this theoretical result for worst case scenarios by showcasing how setting personalized reserve prices with ML advice can safeguard individual welfare in an average sense, using semi-synthetic data derived from real auction data in a search ad platform.

We remark that in this section we only experiment with VCG auctions, particularly because bidders' bidding strategies can be restricted to uniform bidding (see Proposition \ref{prop:ubid} and discussions thereafter), whereas computing near-optimal/best response bidding strategies in GSP and GFP auctions introduces an additional dimension of complexity without adding new insights; see discussions in  \cite{deng2021towards,balseiro2021robust,ai2022no}.

\subsection{Experiment procedure}
\label{subsec:procedure}
\paragraph{Data Generation.}
We derive semi-synthetic data from ad auction logs of a search ad platform with which we simulate VCG auctions for auto bidders. In particular, each entry $j$ of the dataset contains the necessary information to rerun an auction, including the number of ad slots sold $\numslot_{j}$, the CTRs for each ad slot $\bm{\mu}_{j} = (\mu_{j}(1),\dots \mu_{j}(\numslot_{j}))\in [0,1]^{\numslot_{j}}$, a set of candidates and their respective conversion values $v_{i,j}$, etc. Here, the conversion value of an advertiser is the perceived monetary value that she would attain due to a user conversion action (e.g. download, email sign-up, in-app purchase etc).
We emphasize that the main objective for our empirical study
is to validate key insights for our theoretical findings rather than investigating the individual welfare of real systems actually implemented in practice.




\paragraph{Generating ML advice.}
For accuracy level $\beta \in \{ 0.25, 0.5, 0.75\}$, we sample $s_{i,j}^{\beta} \sim \text{Uniform}[\beta,1]$ independently for each $i,j$. We then calculate ML advice $\underline{v}_{i,j}^{\beta} = s_{i,j}^{\beta} \cdot v_{i,j}$, and it is obvious that the ML advice produced this way is $\beta$-accurate, i.e. $\underline{v}_{i,j}^{\beta} \in [\beta v_{i,j},v_{i,j}]$ as defined in Definition \ref{def:reserve}.

\paragraph{Calculating uniform bid multipliers.} 
Here we describe the procedure to generate bid multipliers for bidders in two scenarios, one with reserve prices set with ML advice we generated earlier, and the other without reserve price which we call the ``control experiment''  (for consistency we let $\beta=0$ correspond to this no-reserve price setup). We first describe the procedure to generate bid multipliers for the scenario with reserve prices. In particular, we calculate each advertiser's uniform bid multiplier using gradient descent to emulate uniform bidding practices in reality, because descent/primal-dual methods have been widely adopted for real world autobidding and has proven to have near-optimal convergence and performance guarantees (see e.g. \cite{nesterov2003introductory,balseiro2017budget,aggarwal2019autobidding}. Formally, for each accuracy level $\beta \in\{ 0.25, 0.5, 0.75\}$, we run $2T$ rounds of gradient descent, where in each round we keep the auction environment, including advertiser values, number of slots and CTRs the same  as those we derived from the aforementioned semi-synthetic data. 

The first $T$ rounds are dedicated to ``warm start'' our treatments for reserve prices with different accuracy level: we simulate VCG auctions without reserves until all bidders' uniform bid multipliers convergence. In particular, with an initial uniform bid multiplier $\alpha_{i,1}$ for bidder $i$, for each round $t \in [T]$, set advertiser $i$'s bid multiplier to be $\alpha_{i,t}$, and run $M$ VCG auctions without reserves, where values are $\bm{v} = (v_{i,j})_{i\in [N],j\in [M]}$, ad slot numbers  are $(\numslot_{j})_{j\in [M]}$ and corresponding CTRs are $(\bm{\mu}_{j})_{j\in [M]}$. For round $t\in[T]$, denoting $w_{i,t}$ and $p_{i,t}$ as the total realized welfare and payment for bidder $i$ across all $M$ auctions, we update the uniform bid multiplier with gradient descent in the log-space (note that a similar approach has been used in \cite{deng2021towards,balseiro2021robust})
\begin{align}
    \log \alpha_{i,t+1} =  (1-\eta_{t})\log\alpha_{i,t} +\eta_{t}\log\frac{w_{i,t}}{p_{i,t}}\,. \label{eq:gradient-descent}
\end{align}
Here $\eta_{t}$ are properly chosen step-sizes that ensure convergence within $T$ rounds. Intuitively, when $w_{i,t} > p_{i,t}$, the per-round ROAS balance for bidder $i$ is positive, so in the next round the bidder would have leeway to bid more aggressively with larger bid multipliers to acquire more welfare.
We take the bid-multiplier value $\alpha_{i,T+1}$ for $\beta \in \{0, 0.25, 0.5, 0.75\}$ to be our initial uniform bid multipliers for treatments of VCG auctions with personalized reserves. 

For the next $T$ rounds, we  repeat the above procedure for running VCG auctions, but  with personalized reserve prices $r_{i,j} = \underline{v}_{i,j}^{\beta}$ for treatment trials corresponding to $\beta \in\{ 0.25, 0.5, 0.75\}$, respectively, where we compute $(\alpha_{i,t}^{\beta})_{i\in [N],t = T+1\dots 2T}$ according to Eq. \eqref{eq:gradient-descent}. Finally, we arrive at our bid multipliers $\alpha_{i}^{\beta}  := \alpha_{i, 2T}^{\beta}$.

We repeat the above procedure for our control experiment with no reserve price, and again arrive at bid multipliers $\alpha_{i}^{\beta}  := \alpha_{i, 2T}^{\beta}$ for $\beta = 0$.

\subsection{Experiment results}

For each accuracy level  $\beta \in\{ 0, 0.25, 0.5, 0.75\}$ where we recall $\beta = 0 $ corresponds to the control experiment with no personalized reserve prices, we let $W_{i}^{\beta}$ be the realized total welfare for advertiser $i \in [N]$ across $M$ VCG auctions w.r.t. bid multipliers  $(\alpha_{i}^{\beta})_{i \in [N]}$, values  $\bm{v} = (v_{i,j})_{i\in [N],j\in [M]}$, ad slot numbers   $(\numslot_{j})_{j\in [M]}$ and CTRs $(\bm{\mu}_{j})_{j\in [M]}$ (see details in Subsection \ref{subsec:procedure}). 

\begin{figure*}[ht]
    \centering
\includegraphics[width=0.8\linewidth]{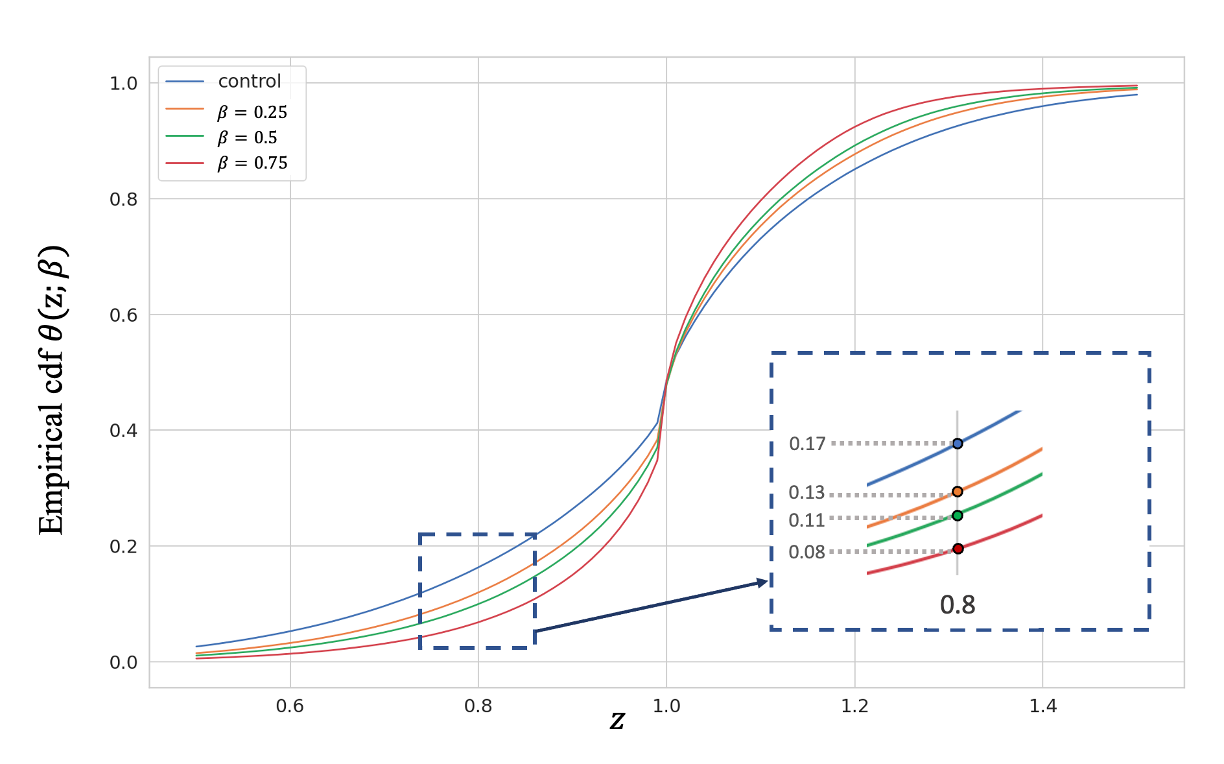}
    \caption{Empirical cdf defined as $\theta(z;\beta) = \frac{1}{N}\sum_{i \in [N]}\I\left\{\frac{W_{i}^{\beta}}{\opt_{i}}\leq z\right\}$ for all accuracy levels $\beta = 0.25, 0.5,0.75$ and the control experiment where we set no personalized reserve prices. 
    For fixed $z <1$, the empirical cdf decreases in accuracy level $\beta$. As an example when $z = 0.8$, $\theta(z= 0.8;\beta = 0) = 0.17$, $\theta(z= 0.8;\beta = 0.25) = 0.13$, $\theta(z= 0.8;\beta = 0.5) = 0.11$, and $\theta(z= 0.8;\beta = 0.75) = 0.08$. }
    \label{fig:empirical}
\end{figure*}

Recalling $\opt_{i}$ as the welfare of bidder $i\in [N]$ under the efficient outcome (see Eq. \eqref{eq:optefficient}), we define the empirical cdf $\theta(z;\beta) = \frac{1}{N}\sum_{i \in [N]}\I\left\{\frac{W_{i}^{\beta}}{\opt_{i}}\leq z\right\}$ to be the proportion of advertisers whose realized welfare is at most least $z$ times her welfare under the efficient outcome. Note that $z$ can be less or greater than 1, where $\theta(z;\beta) $ for $z < 1$ is the proportion of advertisers who incur a welfare loss compared to $\opt_{i}$.

In  Figure \ref{fig:empirical}, we plot the empirical distribution $\theta(z;\beta)$ for $\beta\in\{ 0, 0.25, 0.5, 0.75\}$. We observe that for $z < 1$, $\theta(z;\beta)$ decreases as $\beta$ increases, meaning as we set personalized reserve prices with more accurate ML advice, the proportion of advertisers who lose at least $(1-z) \times 100\%$ of their welfare under the efficient outcome drops. As an example when $z = 0.8$ as illustrated in Figure \ref{fig:empirical} which corresponds to losing at least $20\%$ of the welfare under the efficient outcome, the proportion of bidders who incurs such a welfare loss when $\beta = 0$ (i.e. no reserve prices) is $17\%$, and $13\%$, $11\%$, $8\%$ for $\beta = 0.25$, 0.5, 0.75, respectively. The main takeaway is  improving ML advice accuracy enables the ad platform to decrease the number of advertisers who incur a certain percentage of welfare loss compared to what would have been attained under the efficient outcome. As another observation, we see that the empirical cdf approaches the step function with a jump at $z = 1$, which intuitively means that as ML accuracy improves, more and more advertisers will approach their efficient outcome welfare.


\bibliographystyle{informs2014}
\bibliography{ref}

\newpage
\begin{APPENDICES}
\begin{center}
\vspace{0.8cm}
    \Large Appendices for\\
    \vspace{0.2cm}
    \Large \textbf{Individual Welfare Guarantees in the Autobidding World with Machine-learned Advice}
    \noindent\makebox[\linewidth]{\rule{1\linewidth}{0.9pt}}
\end{center}
\setcounter{page}{1}

\section{Extended Literature Review}
\label{app:extendlitrev}

\textbf{Mechanism design with constrained bidders.} 
This work considers autobidders who 
aim to maximize  welfare while respecting an ROAS constraint that ensures total spend is less than total acquired value, and thus our work relates to the general theme of mechanism design in the presence of constrained agents. The works of  \cite{pai2014optimal,golrezaei2021auction} study revenue-optimal auction design when bidders who maximize quasi-linear utility are constrained by budgets, and return-on-investment (ROI), respectively.  \cite{balseiro2021landscape} study revenue-maximizing auctions for ROI constrained bidders under different objectives and information structures for values and ROI targets. This work differs from these papers as we do not study new auction formats and platform revenue-optimization, but instead presents insights into how to incorporate  ML advice as reserves in classic auctions like VCG, GSP and GFP  can improve individual bidder welfare.

\textbf{Algorithmic bidding/learning under constraints} The behavior of our bidders of interest are governed by their ROAS constraints, and there has been a growing area of works on bidding-algorithm design under similar financial constraints for online advertising markets. \cite{balseiro2017budget} develops theoretical performance guarantees of the budget pacing strategy for bidders with hard budget cap (see more on budget management strategies in \cite{balseiro2021budget}), while \cite{balseiro2020best} presents a more general mirror descent algorithm for online resource allocation problems. \cite{golrezaei2021bidding} present near-optimal bidding algorithms for bidders with both budget and ROI constraints in expectation. Finally, \cite{deng2023multi} study a multi-channel ad procurement problem under the autobidding setup where ad platforms, i.e. channels, autobid on behalf of advertisers, and the work develops algorithms that optimizes advertisers' interaction decisions with channels to maximize conversion. In this work, we do not study the design of bidding algorithms but instead consider worst case outcomes under any feasible bidding profile.

\textbf{Reserve price optimization.}
Reserve price techniques and optimization have been studied for different auction formats and settings.
In the single-shot second price auction setting
\cite{paes2016field, beyhaghi2021improved, derakhshan2022linear} presents different approaches with theoretical performance guarantees to optimize    personalized reserve prices, while \cite{yuan2014empirical} presents an empirical study on the impact of reserve price on the entire auction system for display advertising. For repeated second price auctions, \cite{golrezaei2019dynamic,golrezaei2019incentive,kanoria2020dynamic} dynamically learn reserve prices to maximize cumulative revenue facing strategic agents, where as \cite{feng2021robust} optimize reserve prices to balance revenue and bidders' incentives to misreport. For first price auctions, \cite{feng2021reserve} introduces a gradient-based adaptive algorithm to dynamically optimize reserve prices.  Nevertheless, all aforementioned works attempt to design and learn optimal or near optimal reserve prices for the purpose of revenue maximization, whereas in our work we directly set reserves using ML advice provided by some external black-box, and shed light on how reserve prices can improve individual welfare among all bidders.

\section{Additional materials for Section \ref{sec:model}}
\begin{example}[Personalized-reserve augmented VCG, GSP, GFP auctions]
\label{ex:augmentstandard}
Consider $M\geq 2$ parallel position auctions $(\auc_{j})_{j\in [M]}$ all of which take the form of VCG, GSP or GFP auctions. Each auction $\auc_{j}$ is associated with  $\numslot_{j}\geq 1$ slots and CTRs $\bm{\mu}_{j} = (\mu_{j}(\ell))_{\ell \in \numslot_{j}}$. Assume $N$ bidders submit bid profile $\bm{b}_{j} \in \R_{+}^{N}$ to auction $\auc_{j}$, where $\widetilde{N}_{j}\leq N$ are cleared, i.e. greater than respective personalized reserve prices. Define $\widetilde{\bm{b}}_{j}\in \R_{+}^{\widetilde{N}_{j}}$ to be all ``cleared bids'', and let $\widetilde{b}_{j}^{(\ell)}$ be the $\ell$th highest cleared bid.  
Then, in $\auc_{j}$ bidders who cleared their reserves are assigned slots  according to the ranking of $\widetilde{\bm{b}}_{j}$, whereas the bidders who do not clear their reserves never get allocated any slots. The payment for a bidder $i$ who cleared her reserve and allocated slot $\ell_{i,j} \in [\min\{\widetilde{N}_{j},\numslot_{j}\}]$ is
 \begin{itemize}
    \item \textit{VCG:}  $p_{i,j} = \sum_{\ell = \ell_{i,j}}^{\min\{\widetilde{N}_{j},\numslot_{j}\}}(\mu_{j}(\ell) - \mu_{j}(\ell+1)\cdot \max\{\widetilde{b}_{j}^{(\ell+1)},r_{i,j}\}$ where $\widetilde{b}_{j}^{(\ell)} = 0$ when $\ell > \widetilde{N}_{j}$.
    \item \textit{GSP: }  $p_{i,j} =\mu_{j}(\ell_{i,j})\cdot \max\{\widetilde{b}_{j}^{(\ell_{i,j}+1)},r_{i,j}\}$.
    \item \textit{GFP: } $p_{i,j} =\mu_{j}(\ell_{i,j})\cdot \max\{\widetilde{b}_{j}^{(\ell_{i,j})},r_{i,j}\}$.
\end{itemize}
\end{example}
It is well known that for the same  bid profile $\bm{b}$ and for any bidder $i$,  the payment under the GFP auction is greater than equal to that under GSP auction, and  the payment under the GSP auction is greater than equal to that under VCG; see e.g. \cite{edelman2007internet}. 



\section{Additional materials for Section \ref{sec:VCG}}

\subsection{Proof for Theorem \ref{thm:VCGtight}}
\label{pf:thm:VCGtight}

\begin{theorem}[Restatement of Theorem \ref{thm:VCGtight}]
Consider 2 bidders competing in three SPA auctions whose values are indicated in the following table for any $\beta \in (0,1)$ and ${y}\geq 0$.
\vspace{0.5cm}
\renewcommand{\arraystretch}{1.5}
\begin{center}
\begin{tabular}{|c | c c c| } 
 \hline
  & Auction 1 & Auction 2 & Auction 3 .\\
 \hline\hline
 bidder 1 & ${y}$ & $v$ & 0 \\
 \hline
 bidder 2 & 0  & $v-\epsilon$ & $\gamma + \frac{1}{1-\beta}\cdot \epsilon$  \\
 \hline
\end{tabular}
\end{center}
\vspace{0.5cm}
Bidder 1's multiplier is fixed to be $\alpha_{1} > 1$, and consider $v = \frac{1-\beta}{\alpha_{1}-1} \cdot \gamma$ for any $\gamma > 0$.  The reserve prices are set to be $r_{i,j} = \beta v_{i,j}$. Then, we have
\begin{align}
    \min_{\bm{b}\in \feas}  \frac{\well_{1}(\alloc(\bm{b}))}{\opt_{1}} ~=~ 1- \frac{1-\beta}{\alpha_{1}  -1} \cdot \frac{\opt_{-1} - \frac{1}{1-\beta}\cdot \epsilon}{\opt_{1}}
\end{align}
Taking $\epsilon \to 0$ shows that bidder 1's welfare is equal to the individual welfare guarantee in Theorem \ref{thm:fairnessbound}.
\end{theorem}
\begin{remark}
We remark that as $\epsilon\to 0$, $\frac{\opt_{-i}}{\opt_{i}} = \frac{\frac{\alpha_{1}-1}{1-\beta}v}{{y} + v} \in \left[0, \frac{\alpha_{1}-1}{1-\beta}\right]$, so by varying ${y}\in[0,\infty)$, the above example demonstrates our individual welfare lower bound in Theorem \ref{thm:fairnessbound} is tight for any nontrivial market share ratio  $\frac{\opt_{i}}{\opt_{-i}} \in \Big[\frac{\alpha_{1}-1}{1-\beta},\infty\Big)$.

\end{remark}
\textit{proof}
Note that in any feasible outcome, bidder 1 must win auction 1, and bidder 2 must win auction 3.  Hence for auction 2, we only need to consider the following outcome:

\textbf{Bidder 1 loses auction 2, and suffers welfare loss $v$.} This outcome can be achieved by setting $\alpha_{2}$ such that $\alpha_{2} (v-\epsilon) > \alpha_{1}v$. Bidder 2 accumulates value $v + \gamma + \left( \frac{1}{1-\beta}-1\right)\epsilon$. Her payment for auction 2 is $\max\{\alpha_{1} v ,\beta(v-\epsilon)\}$, and for auction 3 is $\beta\left(\gamma + \frac{1}{1-\beta}\cdot \epsilon\right)$. The following shows that her ROAS constraint is satisfied:
\begin{align*}
    & v + \gamma + \left( \frac{1}{1-\beta}-1\right)\epsilon - \max\{\alpha_{1} v ,\beta(v-\epsilon)\} - \beta\left(\gamma + \frac{1}{1-\beta}\cdot \epsilon\right)\\
    ~\overset{(a)}{=}~ & v + \gamma + \left( \frac{1}{1-\beta}-1\right)\epsilon - \alpha_{1} v  - \beta\left(\gamma + \frac{1}{1-\beta}\cdot \epsilon\right)\\
     ~=~ & \left( 1 - \alpha_{1} \right)v + (1-\beta)\gamma + \left(\frac{1}{1-\beta}-1 - \beta\cdot \frac{1}{1-\beta}\right)\epsilon\\
      ~=~ & 0\,,
\end{align*}
where in (a) we used the fact $\beta \leq 1 < \alpha_{1}$ and $\epsilon  \to 0$. In the final equality we used the definition that $v = \frac{1-\beta}{\alpha_{1}-1} \cdot \gamma$. On the other hand, bidder 1's ROAS constraint is apparently satisfied. 

Under this outcome, denoting bidder 1's welfare as $\well_{1}$ we have 
\begin{align*}
    \frac{\well_{1}}{OPT_{1}} = 1- \frac{v}{OPT_{1}}= 1- \frac{1-\beta}{\alpha_{i} -1} \cdot \frac{\gamma}{\opt_{i}} =  1- \frac{1-\beta}{\alpha_{i} -1} \cdot \frac{\opt_{-i} - \frac{1}{\beta}\cdot \epsilon}{\opt_{i}}
\end{align*}

\halmos

\subsection{Proof of Proposition \ref{prop:boundloss}}
\label{pf:prop:boundloss}

For simplicity, denote $\delta_{i,j}= \opt_{i,j}-\well_{i,j}(\bm{\outcome})$. Then, $\opt_{i}-\well_{i}(\bm{\outcome}) = 
\sum_{j\in [M]:\delta_{i,j}>0}\delta_{i,j}  + \sum_{j\in [M]:\delta_{i,j}=0}\delta_{i,j} + \sum_{j\in [M]:\delta_{i,j}<0}\delta_{i,j} =
\loss_{i}(\bm{\outcome}) + \sum_{j\in [M]:\well_{i,j}(\bm{\outcome}) > \opt_{i,j}}\left(\opt_{i,j}-\well_{i,j}(\bm{\outcome})\right) \leq \loss_{i}(\bm{\outcome}) \leq B\,.$ Rearranging and dividing both sides by $\opt_{i}$ we get $\frac{\well_{i}(\bm{\outcome})}{\opt_{i}}\geq 1- \frac{B}{\opt_{i}}.$ 

Here we remark that it is possible to have $\well_{i,j}(\bm{\outcome}) > \opt_{i,j}$ because bidders may overbid, and therefore win auctions/slots that they would not have won under the efficient outcome.
\halmos

\subsection{Proof of Lemma \ref{lem:validmultiregion}}
\label{pf:lem:validmultiregion}

Recall there is $\Delta$-separation in values. Fix a bidder $i$ and let $v_{j}^{+}$ be the smallest competitor value  that is strictly greater than $v_{i,j}$ in any auction $\auc_{j}$ where bidder $i$'s value is not the largest, and by definition of $\Delta$-separated values we have $v_{j}^{+} \geq \Delta v_{i,j}$. Hence, by using any multiplier $\alpha_{i} \in [1,\Delta)$ and assuming competitors bid truthfully, the outcome of the auctions would be identical to that of everyone (including bidder $i$) bidding truthfully. And since truthful bidding is always feasible, we conclude that $\bm{v}_{-i} \in \feas_{-i}^{u}(\alpha_{i}\bm{v}_{i})$ for $\alpha_{i} \in [1,\Delta)$.

\halmos

\section{Additional materials for Section \ref{sec:imposs}}

\subsection{Additional Definitions and Lemmas for Section \ref{sec:imposs}}
\label{app:imposs:add}
The following lemma shows that for anonymous and truthful auctions, the probability of the lowest bidder winning a single auction is capped by a bound that decreases as the number of bidders grow.
\begin{lemma}[Lemma 3 in \cite{mehta2022auction}]
\label{lem:winprobub}
In an anonymous and truthful auction for a single item with $N$ bidders, the bidder who submits the lowest bid wins the item with probability at most $\frac{1}{N}$.
\end{lemma}

The following technical definition and lemma (i.e. Definition \ref{def:monop} and Lemma \ref{lem:singlebiddercost}) concerns the scenario where only one bidder participates in the auction (others bid 0), and present an upper bound on the probability and cost  respectively for the single bidder to win the auction.
\begin{definition}[Single bidder purchase probability and bid threshold]
\label{def:monop}
For any allocation-anonymous and truthful auction $\mathcal{A}$, consider the setting with a single bidder who submits bid $b > 0$ and define
\begin{align}
    \pi_{\mathcal{A}} = \lim_{b\to \infty}\prob(\text{bidder wins item with bid $b$})\,,
\end{align}
where the limit exists because in a truthful auction, $\prob(\text{bidder wins item with bid $b$})$ increases in $b$ (see Definition \ref{def:truthfulauctions} for truthful auctions). Assume this max probability is reached at some bid threshold $Q_{\mathcal{A}}$, i.e.
\begin{align}
   Q_{\mathcal{A}} = \min\left\{b>0: \prob(\text{bidder wins item with bid $b$}) =  \pi_{\mathcal{A}}\right\} \,.
\end{align}
\end{definition}
Note that in a deterministic single-slot auction that allocates to the highest bidder,  $\pi_{\mathcal{A}} = 1$, and $Q_{\mathcal{A}}  \to 0$. For example, in an SPA  with no reserve, the single bidder can win the auction with any arbitrarily small positive bid with probability 1.

\begin{lemma}[Lemma 4 in \cite{mehta2022auction}]
\label{lem:singlebiddercost}
For any allocation-anonymous and truthful auction $\mathcal{A}$ with single-bidder purchase probability $ \pi_{\mathcal{A}}$ and bid threshold $ Q_{\mathcal{A}}$, the expected cost for a single bidder for winning the item is at most $\pi_{\mathcal{A}} \cdot Q_{\mathcal{A}}$.
\end{lemma}

\subsection{Proof of Theorem \ref{thm:imposs}}
\label{pf:thm:imposs}
\begin{theorem}[Restatement of Theorem \ref{thm:imposs}]
For any auction $\mathcal{A}$ that is allocation-anonymous, truthful, and possibly randomized, \footnote{Here, we assume all auctions of interest are individually rational (IR), i.e. the payment of a bidder is always less than her submitted bid.} consider an autobidding problem instance w.r.t.  $\mathcal{A}$ with $M = 2K+1$ auctions and $N = K+1$ bidders. Fix bidder 0's bid multiplier to be $\alpha_{0}$ and some $\beta\in [0,1)$. Consider the bidder values $\{v_{i,j}\}_{i\in[N],j\in[M]}$ given in the following table. 
\vspace{1cm}
\renewcommand{\arraystretch}{2}
\begin{center}
\begin{tabular}{|c | c c c c c c c c c|} 
 \hline
  & $A_{1}$ & $A_{2}$ & \dots  &   $A_{K}$  & $A_{K+1}$ & $A_{K+2}$ & \dots & $A_{2K}$ &  $A_{2K+1}$\\
 \hline\hline
$B_{1}$  & $\frac{\alpha_{0}v+\epsilon}{\rho}$ & $\frac{\alpha_{0}v+2\epsilon}{\rho}$  & \dots &  $\frac{\alpha_{0}v+K\epsilon}{\rho}$ & $\gamma$  & 0 &  \dots & 0 & 0\\
  \hline
$B_{2}$   & $\frac{\alpha_{0}v+2\epsilon}{\rho}$ & $\frac{\alpha_{0}v+3\epsilon}{\rho}$  & \dots &  $\frac{\alpha_{0}v+\epsilon}{\rho}$ & 0 &  $\gamma$ & \dots &  0 & 0  \\
 \hline
 \vdots & \vdots &  \vdots  &  &   \vdots  &  \vdots & \vdots & & \vdots & \vdots \\
  \hline
$B_{K}$  & $\frac{\alpha_{0}v+K\epsilon}{\rho}$ & $\frac{\alpha_{0}v+\epsilon}{\rho}$  & \dots &  $\frac{\alpha_{0}v+(K-1)\epsilon}{\rho}$ & 0  & 0  & \dots & $\gamma$ & 0\\
   \hline
$B_{0}$  & $v$ & $v$ & \dots & $v$ & 0 & 0 & \dots & 0  & ${y}$\\ 
 \hline
\end{tabular}
\end{center}
\vspace{1cm}
In the table, we let $\gamma > \frac{Q_{\mathcal{A}}}{\beta} > Q_{\mathcal{A}}$, $\epsilon = O(1/K^{3})$ and $v = \frac{1-\beta}{\alpha_{0}-1}\cdot \pi_{\mathcal{A}}\cdot \gamma$. Let $\rho,{y}$ and a large enough $K$ satisfy the following:
\begin{align}
\begin{aligned}
\label{eq:imposs:paramsnew}
   & \alpha_{0}<\rho <\frac{\alpha_{0}}{\beta} ~\text{s.t.}~\frac{\alpha_{0}v+K\epsilon}{\rho} < v,\quad \text{ and } {y} > \max\left\{\frac{Q_{\mathcal{A}}}{\alpha_{0}}, \frac{\alpha_{0}v}{\pi_{\mathcal{A}}}\right\}\,,
  \end{aligned}
\end{align}

where $Q_{\mathcal{A}},\pi_{\mathcal{A}}$ are defined in Definition \ref{def:monop}. Further, suppose the platform enforces personalized reference prices $\bm{r}\in \R_{+}^{N\times M}$ on top of auction $\mathcal{A}$, where $r_{i,j} = \beta v_{i,j}$.
 Then, letting the (possibly random) outcome be $\bm{x}$ when bidders 1, ... K all adopt the bid multiplier $\rho$, 
 the ROAS constraints for all bidders are satisfied when $K\to\infty$ and $\rho \to \alpha_{0}$, and for bidder $0$ we have 
 \begin{align}
     \lim_{K\to \infty}\frac{\E_{\mathcal{A}}\left[\well_{0}(\bm{x})\right]}{\E_{\mathcal{A}}\left[\opt_{0}\right]} \leq  1-  \frac{1-\beta}{\alpha_{0}-1} \cdot \lim_{K\to \infty}\frac{\E_{\mathcal{A}}\left[\opt_{-0}\right]}{\E_{\mathcal{A}}\left[\opt_{0}\right]}
 \end{align}
 where $\E_{\mathcal{A}}$ is taken w.r.t. the randomness in outcome $\bm{x}$ due to randomness in the auction $\mathcal{A}$.
\end{theorem}

\textit{Proof.}
First note that bidder 0 only has competition in auctions $A_{1}...A_{K}$, and hence can only incur a loss (that contributes to $\loss_{0}(\bm{x})$ defined in  Equations (\ref{eq:loss})) within these auctions. Hence $ \E_{\mathcal{A}}[\loss_{0}(\bm{x})]  =  v\sum_{j\in [K]}\prob(\text{bidder 0 loses auction $j$})$. Then we consider the following:
\begin{align}
\begin{aligned}
    \E_{\mathcal{A}}[\loss_{0}(\bm{x})] ~=~ &  v\sum_{j\in [K]}\prob(\text{bidder 0 loses auction $j$})~=~  v\sum_{j\in [K]}\left(1-\prob(\text{bidder 0 wins auction $j$})\right)\\
    ~\overset{(a)}{\geq}~ & v \cdot \frac{K^{2}}{K +1} = \frac{1-\beta}{\alpha_{0}-1}\cdot \gamma \cdot \pi_{\mathcal{A}} \cdot \frac{K^{2}}{K +1} \overset{(b)}{=}  \frac{1-\beta}{\alpha_{0}-1} \cdot \E_{\mathcal{A}}\left[\opt_{-0}\right] \cdot \frac{K}{K+1}\,.
    \end{aligned}
\end{align}
Here (a) holds because bidder 0 bids $\alpha_{0}v$ for any auction in 1,2...K, which is strictly less than all other bidders' bids as they all adopt  multipliers $\rho$ in these auctions, so from Lemma \ref{lem:winprobub}, we have $\prob(\text{bidder 0 wins auction $j$})\leq \frac{1}{K+1}$; in (b) we used the fact that $\E_{\mathcal{A}}\left[\opt_{-0} \right]= \sum_{j=K+1}^{2K}\E_{\mathcal{A}}\left[\gamma\right] = \gamma \cdot K \cdot \pi_{\mathcal{A}}$ since there is only a single non-zero bidder in auctions $A_{K+1}\ldots A_{2K}$ and each bidder submits a bid $\rho \gamma > \rho > Q_{\mathcal{A}}$ (see Definition \ref{def:monop}).

Therefore we have 

\begin{align}
    \lim_{K\to \infty}\frac{\E_{\mathcal{A}}\left[\well_{0}(\bm{x})\right]}{\E_{\mathcal{A}}\left[\opt_{0}\right]} \overset{(a)}{=}  1 - \lim_{K\to \infty}\frac{\E_{\mathcal{A}}\left[\loss_{0}(\bm{x})\right]}{\E_{\mathcal{A}}\left[\opt_{0}\right]}
     \leq 1-  \frac{1-\beta}{\alpha_{0}-1} \lim_{K\to \infty}\frac{\E_{\mathcal{A}}\left[\opt_{-0}\right]}{\E_{\mathcal{A}}\left[\opt_{0}\right]}\,,
\end{align}
where (a) follows from the fact that  in our constructed autobidding instance, bidder {0}'s acquired value in each auction cannot exceed that under the efficient allocation, and hence can only incur loss in welfare.

Now it only remains to show that the multiplies $(\alpha_{0}, \rho, \dots \rho) \in (1,\infty)^{K+1}$ yields a feasible outcome, i.e. the ROI constraints of each bidder is satisfied in expectation. Let $V_{i,j}$ and $C_{i,j}$  be the expected value and cost of bidder $i$ in auction $A_{j}$, respectively.

\textbf{1. Showing bidder $0$'s ROI constraint is satisfied}.
We show by the following: bidder 0 only incurs a non-zero expected cost in auctions $A_{1}\dots A_{K}$ and $A_{2K+1}$, and we will show that the expected value $V_{0,2K+1}$ is lower bounded by the expected costs $C_{0,2K+1} + \sum_{j\in[K]}C_{0,j}$.

Since $\alpha_{0}{y}  > Q_{\mathcal{A}}$, the definition of the single-bidder purchasing probability in Definition \ref{def:monop} implies that  bidder 0 acquires an expected value from  auction $A_{2K+1}$ of $V_{0,2K+1} = \pi_{\mathcal{A}}{y}$. Further,  since bidder 0 is submits the lowest bids in auctions $A_{1} \dots A_{K}$ under bid multiplier profile $(\alpha_{0},\rho \dots \rho)\in (0,\infty)^{K+1}$,  from Lemma \ref{lem:winprobub}, we have $\prob(\text{bidder 0 wins auction $j$})\leq \frac{1}{K+1}$ for all $j \in [K]$. Since the payment of a bidder in an auction is at most her submitted bid (as the auction is IR), we know that $\sum_{j \in [K]}C_{0,j}\leq  K\cdot \frac{\alpha_{0}v}{K+1} < \pi_{\mathcal{A}}{y} = V_{0,2K+1}$, where the inequality follows from the definition of ${y}$ in Equation (\ref{eq:imposs:paramsnew}) such that ${y} > \max\left\{\frac{Q_{\mathcal{A}}}{\alpha_{0}}, \frac{\alpha_{0}v}{\pi_{\mathcal{A}}}\right\}$. This implies bidder 0's ROI constraint is satisfied.

\textbf{2. Showing bidder $i$'s ROI constraint is satisfied for any $i = 1,2 \dots K$}. We show this by considering the following: bidder $i$ only incurs a non-zero expected cost in auctions $A_{1}\dots A_{K}$ and $A_{K+i}$, and we will show that the expected values $ V_{i,K+i} + \sum_{k\in [K]}V_{i,j}$ is lower bounded by the expected costs $C_{i,K+i} + \sum_{j\in[K]}C_{i,j}$.
\begin{itemize}
 
    \item \textbf{Calculate cost $C_{i,K+i}$: } For auction $A_{K+i}$, bidder $i$'s bid is $\rho\gamma > \gamma > Q_{\mathcal{A}}$ from the definition of $\gamma$, so by Definition \ref{def:monop}, the probability of $i$ winning the item in auction $A_{K+i}$ is $\pi_{\mathcal{A}}$, and the expected cost is
    \begin{align}
    \label{eq:imposs:costub}
     C_{i,K+i} \leq \pi_{\mathcal{A}}\cdot \max\left\{r_{i,K+i}, Q_{\mathcal{A}}  \right\} \leq \pi_{\mathcal{A}}\cdot\beta \gamma\,,
     \end{align}
     where the final inequality follows from the definition $r_{i,K+i} = \beta \gamma  $
     
     \item \textbf{Upper bound costs $\sum_{j\in[K]}C_{i,j}$: }  
             For auctions $[K] = 1\dots K$,  bidder $i$'s total expected cost can be bounded as
        \begin{align}
        \begin{aligned}
        \label{eq:imposs:costub2}
           \sum_{j\in[K]} C_{i,j} ~\leq~ & \rho \sum_{j\in[K]} v_{i,j} \prob\left(\text{bidder $i$ wins auction $A_{j}$}\right)\\
           ~=~ & \alpha_{0}v \sum_{j \in [K]}\prob\left(\text{bidder $i$ wins auction  $A_{j}$}\right) + \frac{(K+1)K}{2}\epsilon \,.
          \end{aligned}
        \end{align}
    where the first inequality follows from a bidder's payment is at most her submitted bid since the auction is IR.
     \item \textbf{Calculate $V_{i,K+i}$: }
        Considering auction $A_{K+i}$, bidder $i$ is the only bidder, and since $\rho \gamma >\gamma > Q_{\mathcal{A}}$, the definition of the single-bidder purchasing probability in Definition \ref{def:monop} implies that  bidder $i$'s acquires an expected value from this auction of 
    \begin{align}
    \label{eq:imposs:vallb}
        V_{i,K+i} = \pi_{\mathcal{A}}\cdot \gamma\,.
    \end{align} 
    
    \item \textbf{Lower bound $\sum_{k\in [K]}V_{i,j}$: }
    
    \begin{align}
    \label{eq:imposs:vallb2}
        \sum_{k\in [K]}V_{i,j} \geq \frac{\alpha_{0}v}{\rho}\sum_{j\in [K]}\prob\left(\text{bidder $i$ wins auction $j$}\right)\,.
    \end{align}
\end{itemize}

Combining Equations  (\ref{eq:imposs:costub}),(\ref{eq:imposs:costub2}),(\ref{eq:imposs:vallb}) and (\ref{eq:imposs:vallb2}), we get
\begin{align}
\label{eq:imposs:vallb3}
\begin{aligned}
     & \sum_{j\in[K]}V_{i,j}  + V_{i,K+i} -  \left(\sum_{j\in[K]}C_{i,j}+ C_{i,K+i}\right) \\ 
     ~\geq~ & \pi_{\mathcal{A}}\cdot \gamma  +\frac{\alpha_{0}v}{\rho} \cdot \sum_{j\in [K]}\prob\left(\text{bidder $i$ wins auction $j$}\right) \\
     &\quad - \left(\pi_{\mathcal{A}}\cdot\beta \gamma + \alpha_{0}v\cdot \sum_{j\in [K]}\prob\left(\text{bidder $i$ wins auction $j$}\right) +  \frac{(K+1)K}{2}\epsilon\right)\\
     ~=~ &  \pi_{\mathcal{A}}\cdot (1-\beta) \gamma - \left(\alpha_{0}- \frac{\alpha_{0}}{\rho}\right) v \cdot \sum_{j\in [K]}\prob\left(\text{bidder $i$ wins auction $j$}\right) -  \frac{(K+1)K}{2}\epsilon\\
     ~\overset{(a)}{=}~ &  (\alpha_{0}-1)v - \left(\alpha_{0}- \frac{\alpha_{0}}{\rho}\right) v \cdot \sum_{j\in [K]}\prob\left(\text{bidder $i$ wins auction $j$}\right)-  \frac{(K+1)K}{2}\epsilon\\
      ~\overset{(b)}{\geq}~ &  (\alpha_{0}-1)v - \left(\alpha_{0}- \frac{\alpha_{0}}{\rho}\right) v -  \frac{(K+1)K}{2}\epsilon\\
     \,,
    \end{aligned}
\end{align}
where (a) follows from the definition $v = \frac{1-\beta}{\alpha_{0}-1}\cdot \pi_{\mathcal{A}}\cdot \gamma$; In (b) we used the fact that $\rho > \alpha_{0}>1$ and $ \sum_{j \in [K]}\prob\left(\text{bidder $i$ wins auction  $A_{j}$}\right)\leq 1 $ due to the following:
Consider the set of bid values $\mathcal{B} = \{\alpha_{0}v, \alpha_{0}v+\epsilon, \alpha_{0}v+2\epsilon \dots \alpha_{0}v+ K\epsilon\} \subseteq \R_{> 0}$, and  we recognize that any bid value $b_{k} \in \mathcal{B}$ exceeds the maximim reserve price $\beta v$ in auctions $A_{1} ... A_{K}$. Therefore the constructed reserve prices do not affect allocation, and hence by anonymity of auction $\mathcal{A}$ there exists probabilities $\bm{q}(\mathcal{B})=(q_{0}(\mathcal{B}),q_{1}(\mathcal{B}) \dots q_{K}(\mathcal{B})) \in [0,1]^{K+1}$ where $$q_{k}(\mathcal{B}) = \prob(\text{bid value $b_{k}$ wins auction $\mathcal{A}$ given competing bids $\bm{b}_{-k}$}) \quad \text{ and } \sum_{k=0}^{K}q_{k}(\mathcal{B})\leq 1.$$
We recognize that  in each auction $A_{1} \dots A_{K}$,  under bid multipliers $(\alpha_{0},\rho,\dots \rho)\in (1,\infty)^{K+1}$ the submitted bid profile is a cyclic permutation of  $\mathcal{B}$. Therefore we know that
\begin{align*}
    \sum_{j \in [K]}\prob\left(\text{bidder $i$ wins auction $j$}\right) =  \sum_{k = 1}^{K} q_{k}(\mathcal{B}) \leq 1-  q_{0}(\mathcal{B})\leq 1
\end{align*}

Finally, by taking $\rho \to \alpha_{0}$ and $K\to\infty$ in Equation (\ref{eq:imposs:vallb3}), and utilizing $\epsilon = O(1/K^{3})$ we have
\begin{align*}
& \lim_{\rho \to \alpha_{0}} \lim_{K\to\infty}\sum_{j\in[K]}V_{i,j}  + V_{i,K+i} -  \left(\sum_{j\in[K]}C_{i,j} + C_{i,K+i}\right)~\geq~ 0 \,.
\end{align*}
This shows that bidder $i$'s ROI constraint is satisfied.

\halmos

\section{Proofs for Section \ref{sec:extensions}}
\subsection{Proof of Theorem \ref{thm:GFP:fairnessbound}}
\label{pf:thm:GFP:fairnessbound}
For convenience, define $\delta = 2-\frac{1}{\Delta} $, so $\Delta > 1$ implies $\delta \in (1,2)$, and further $1>\beta > \frac{\Delta}{2\Delta-1}$ implies $\frac{1}{\delta} < \beta < 1$.

Fix a bidder $i \in [K]$ and  any feasible competing bid profile $\bm{b} \in \mathcal{U}$. Denote the corresponding outcome as $\bm{\outcome} = \alloc(\bm{b})$, where $\bm{\outcome}=(\bm{\outcome}_{1} ... \bm{\outcome}_{M})$ where $\bm{\outcome}_{j} \in \{0,1\}^{N\times \numslot_{j}}$ is the outcome vector in auction $\auc_{j}$. Note that by definition of $\mathcal{U}$ which is the set of undominated and feasible bids, under the outcome $\bm{\outcome}$ all bidders' ROAS constraints are satisfied. Denote $\ell_{k,j}$, $\ell_{k,j}^{*}$ to be the position of bidder $k\in[N]$ in auction $j\in [M]$ under outcome $\bm{x}$ and the efficient outcome, respectively.

Recall in Eq.\eqref{eq:covering} the definition for the set of all ``coverings'' for bidder $i$, denoted as $\mathcal{C}_{i}(\bm{\outcome})$:
\begin{align*}
\begin{aligned}
   & \mathcal{B}_{i}(k;\bm{x})= \left\{j\in [M]: \opt_{i,j} > 0,~ v_{k,j} < v_{i,j} \text{ and } \ell_{k,j} \leq \ell_{i,j}^{*} <  \ell_{i,j}\right\}\\
   &  \mathcal{C}_{i}(\bm{\outcome}) = \left\{\mathcal{C} \subseteq [N]/\{i\}: (\mathcal{B}_{i}(k;\bm{x}))_{k\in \mathcal{C}} \text{ is a maximal set cover of }\mathcal{L}_{i}(\bm{\outcome})\right\}\,
    \end{aligned}
\end{align*}
where $\mathcal{L}_{i}(\bm{\outcome}) = \{j\in [M]: \well_{i,j}(\bm{\outcome})< \opt_{i,j}\}$ is the set of auctions in which bidder $i$'s acquired welfare is less than that of her welfare under the efficient outcome; see Definition \ref{def:welfareloss}.

Denote $\pay_{k,j}$ as the payment of any bidder $k$, and
$\Hat{b}_{\ell,j}$ as the $\ell$th largest bid in any auction $j\in[M]$. Similar to the proof of Theorem 
\ref{thm:fairnessbound}, fix any covering $\mathcal{C} \subseteq \mathcal{C}_{i}(\bm{\outcome})$, and any bidder $k \in \mathcal{C}$, such that in some auction $j\in \mathcal{B}_{i}(k;\bm{x})$, we have $v_{k,j}<v_{i,j}$ but $ \ell_{k,j} \leq \ell_{i,j}^{*} <  \ell_{i,j}$. Thus following a similar deduction as Eq. \eqref{eq:VCG:badbidder1} in the proof of Theorem \ref{thm:fairnessbound}, bidder $k$'s payment is lower bounded as
\begin{align}
\begin{aligned}
\label{eq:GFP:badbidder1}
   \text{For }j \in \mathcal{B}_{i}(k;\bm{x}),\quad p_{{k},j}
    ~\overset{(a)}{\geq}~& \sum_{\ell = \ell_{{k},j}}^{\numslot_{j}}\left(\mu(\ell) - \mu(\ell+1)\right)\Hat{b}_{\ell+1,j}
    \\
    ~=~ & \sum_{\ell = \ell_{{k},j}}^{\ell_{i,j}-1}\left(\mu(\ell) - \mu(\ell+1)\right)\Hat{b}_{\ell+1,j}
     +  p_{i,j}\\
     ~\overset{(b)}{\geq}~&  \beta \left(\mu(\ell_{{k},j}) - \mu(\ell_{i,j})\right)v_{i,j} + \beta \cdot  \mu(\ell_{i,j})v_{i,j}\\
      ~=~&  \beta \mu(\ell_{{k},j})\cdot  v_{i,j}\\
      ~=~&  \mu(\ell_{{k},j}) v_{i,j} + \left(\beta-\frac{1}{\delta}\right)\left(\mu(\ell_{i,j}^{*}) - \mu(\ell_{i,j})\right) v_{i,j} - (1- \beta) \cdot  \mu(\ell_{i,j})v_{i,j}\\
      & -(1-\beta)\mu(\ell_{{k},j}) v_{i,j} + \left(\frac{1}{\delta}-\beta\right)\mu(\ell_{i,j}^{*})  v_{i,j} + \left(1-\frac{1}{\delta}\right)\mu(\ell_{i,j})v_{i,j}\\
      ~\overset{(c)}{\geq}~&  \mu(\ell_{{k},j}) v_{i,j} + \left(\beta-\frac{1}{\delta}\right)\left(\mu(\ell_{i,j}^{*}) - \mu(\ell_{i,j})\right) v_{i,j} - (1- \beta) \cdot  \mu(\ell_{i,j})v_{i,j}\\
      & -\left(1-\frac{1}{\delta}\right)\mu(\ell_{{k},j}) v_{i,j}  + \left(1-\frac{1}{\delta}\right)\mu(\ell_{i,j})v_{i,j}\\
       ~=~&   \left(\beta-\frac{1}{\delta}\right)\left(\mu(\ell_{i,j}^{*}) - \mu(\ell_{i,j})\right) v_{i,j} - (1- \beta) \cdot  \mu(\ell_{i,j})v_{i,j} \\
       & +  \frac{1}{\delta}\mu(\ell_{{k},j}) v_{i,j} +\left(1-\frac{1}{\delta}\right)\mu(\ell_{i,j})v_{i,j}
    \end{aligned}
\end{align}
Here , (a) follows from the fact that for a fix bid profile, the payment of GSP or GFP for each bidder in an auction dominates that of VCG (see Example \ref{ex:augmentstandard} and discussions thereof); (b) follows from $\Hat{b}_{\ell,j} \geq b_{i,j}$ for $\ell\leq \ell_{i,j}$, and since
$\bm{b} \in \mathcal{U}\subseteq\R_{+}^{N\times M}$ is an undominated bid profile, Lemma \ref{lem:undominatedbids} applies and $ b_{i,j}\geq \beta v_{i,j}$. Also $p_{i,j}\geq r_{i,j}\geq \beta v_{i,j}$ be the definition of $\beta$-approximate reserves; (c) follows from the fact that $\beta >\frac{1}{\delta}$ and $\mu(\ell_{i,j}^{*})\leq \mu(\ell_{{k},j})$  since $\ell_{{k},j}\leq \ell_{i,j}^{*}$ for any $k \in \mathcal{C}\subseteq \mathcal{C}_{i}(\bm{\outcome})$ and $j \in \mathcal{B}_{i}(k;\bm{x})$; see definition in Eq. \eqref{eq:covering}.

On the other hand, we have
\begin{align*}
      & \sum_{j\in \mathcal{B}_{i}(k;\bm{x})}p_{{k},j} + \sum_{j\notin \mathcal{B}_{i}(k;\bm{x})}  p_{{k},j} \leq \sum_{j\in \mathcal{B}_{i}(k;\bm{x})}\mu(\ell_{{k},j})v_{{k},j} + \sum_{j\notin \mathcal{B}_{i}(k;\bm{x})}  \mu(\ell_{{k},j})v_{{k},j}\\   
      &  p_{{k},j}\geq \beta\cdot \mu(\ell_{{k},j})v_{{k},j} \quad \forall j\in[M]\,,
\end{align*}
where the first inequality follows from bidder ${k}$'s ROAS constraint; the second  inequality follows from the fact that any winning bidder's payment must be greater than her $\beta$-approximate reserves.

Combining the above inequalities and rearranging we get
\begin{align}
\label{eq:GFP:badbidder2}
   \sum_{j\in \mathcal{B}_{i}(k;\bm{x})} p_{{k},j} \leq \sum_{j\in \mathcal{B}_{i}(k;\bm{x})}\mu(\ell_{{k},j})v_{{k},j} + (1-\beta)\cdot \sum_{j\notin \mathcal{B}_{i}(k;\bm{x})}  \mu(\ell_{{k},j})v_{{k},j}\,,
\end{align}
Summing Eq.\eqref{eq:GFP:badbidder1} over all $j\in \mathcal{B}_{i}(k;\bm{x})$ and combining with Eq. \eqref{eq:GFP:badbidder2}, we get
\begin{align}
\begin{aligned}
\label{eq:GFPbounssingleloss}
  & \left(\beta - \frac{1}{\delta}\right)\cdot \sum_{j\in \mathcal{B}_{i}(k;\bm{x})}\left(\mu(\ell_{i,j}^{*}) - \mu(\ell_{i,j})\right) v_{i,j} \\
  ~\leq~ & (1- \beta) \cdot \left( \sum_{j\in \mathcal{B}_{i}(k;\bm{x})}\mu(\ell_{i,j})v_{i,j} + \sum_{j\notin \mathcal{B}_{i}(k;\bm{x})} \mu(\ell_{{k},j})v_{{k},j}\right) \\
  &\quad + \underbrace{ \sum_{j\in \mathcal{B}_{i}(k;\bm{x})}\mu(\ell_{{k},j})v_{{k},j} - \frac{1}{\delta}\sum_{j\in \mathcal{B}_{i}(k;\bm{x})}\mu(\ell_{{k},j}) v_{i,j} -\left(1-\frac{1}{\delta}\right)\sum_{j\in \mathcal{B}_{i}(k;\bm{x})}\mu(\ell_{i,j})v_{i,j}}_{Y}\,.
  \end{aligned}
\end{align}

We now upper bound $Y$:
\begin{align}
\begin{aligned}
\label{eq:GFPbounssingleloss1}
   & \sum_{j\in \mathcal{B}_{i}(k;\bm{x})}\mu(\ell_{{k},j})v_{{k},j} - \frac{1}{\delta} \sum_{j\in \mathcal{B}_{i}(k;\bm{x})}\mu(\ell_{{k},j}) v_{i,j} -\left(1-\frac{1}{\delta}\right) \sum_{j\in \mathcal{B}_{i}(k;\bm{x})}\mu(\ell_{i,j})v_{i,j}\\
   ~=~ &\left(1-\frac{1}{\delta}\right) \sum_{j\in \mathcal{B}_{i}(k;\bm{x})}\mu(\ell_{{k},j})v_{{k},j} - \frac{1}{\delta} \sum_{j\in \mathcal{B}_{i}(k;\bm{x})}\mu(\ell_{{k},j}) \left(v_{i,j}-v_{{k},j}\right) -\left(1-\frac{1}{\delta}\right) \sum_{j\in \mathcal{B}_{i}(k;\bm{x})}\mu(\ell_{i,j})v_{i,j}\\
    ~=~ &\left(1-\frac{1}{\delta}\right) \sum_{j\in \mathcal{B}_{i}(k;\bm{x})}\mu(\ell_{{k},j})\left(v_{{k},j} - v_{i,j}\right) - \frac{1}{\delta} \sum_{j\in \mathcal{B}_{i}(k;\bm{x})}\mu(\ell_{{k},j}) \left(v_{i,j}-v_{{k},j}\right)\\
    & + \left(1-\frac{1}{\delta}\right) \sum_{j\in \mathcal{B}_{i}(k;\bm{x})}\left(\mu(\ell_{{k},j}) - \mu(\ell_{i,j})\right)v_{i,j}\\
     ~=~ &\left(1-\frac{1}{\delta}\right) \sum_{j\in \mathcal{B}_{i}(k;\bm{x})}\mu(\ell_{{k},j})\left(v_{{k},j} - v_{i,j}\right) +  \sum_{j\in \mathcal{B}_{i}(k;\bm{x})}\frac{\mu(\ell_{{k},j})v_{i,j}}{\delta}\left(\left(\delta-1\right)\left(1- \frac{\mu(\ell_{i,j})}{\mu(\ell_{{k},j})} \right) -  \left(1-\frac{v_{{k},j}}{v_{i,j}}\right)\right)\\
    ~\overset{(a)}{\leq}~ &\left(1-\frac{1}{\delta}\right) \sum_{j\in \mathcal{B}_{i}(k;\bm{x})}\mu(\ell_{{k},j})\left(v_{{k},j} - v_{i,j}\right) +  \sum_{j\in \mathcal{B}_{i}(k;\bm{x})}\frac{\mu(\ell_{{k},j})v_{i,j}}{\delta}\left( \left(\delta-1\right) - \left(1-\frac{v_{{k},j}}{v_{i,j}}\right)\right)\\
     ~=~ &\left(1-\frac{1}{\delta}\right) \sum_{j\in \mathcal{B}_{i}(k;\bm{x})}\mu(\ell_{{k},j})\left(v_{{k},j} - v_{i,j}\right) +  \sum_{j\in \mathcal{B}_{i}(k;\bm{x})}\frac{\mu(\ell_{{k},j})v_{i,j}}{\delta}\left( \left(\delta-2\right) +\frac{v_{{k},j}}{v_{i,j}}\right)\\
      ~\overset{(b)}{\leq}~&\left(1-\frac{1}{\delta}\right) \sum_{j\in \mathcal{B}_{i}(k;\bm{x})}\mu(\ell_{{k},j})\left(v_{{k},j} - v_{i,j}\right) \\
      ~\overset{(c)}{\leq}~&\left(1-\beta\right) \sum_{j\in \mathcal{B}_{i}(k;\bm{x})}\mu(\ell_{{k},j})\left(v_{{k},j} - v_{i,j}\right) \,.
   \end{aligned}
\end{align}
where in (a) we recall $\delta > 1$ and $\ell_{{k},j} < \ell_{i,j}$ for any $k\in \mathcal{C}$ and $j\in \mathcal{B}_{i}(k;\bm{x})$ so that $\mu(\ell_{{k},j}) > \mu(\ell_{i,j})$; (b) follows from the fact that values are $\delta$-separated, so  $v_{i,j} > v_{{k},j}$ for $k\in \mathcal{C}$ and $j\in \mathcal{B}_{i}(k;\bm{x})$ implies $\frac{v_{{k},j}}{v_{i,j}}\leq \frac{1}{\Delta}= 2-\delta$; in (c) we used the fact that $\beta > \frac{1}{\delta}$ so $1-\beta < 1-\frac{1}{\delta}$, and the fact that $  v_{{k},j} < v_{i,j}$  for any $k\in \mathcal{C}$ and $j\in \mathcal{B}_{i}(k;\bm{x})$.

Combining Equations (\ref{eq:GFPbounssingleloss}) and (\ref{eq:GFPbounssingleloss1}) we get
\begin{align}
\begin{aligned}
& \left(\beta - \frac{1}{\delta}\right)\cdot \sum_{j\in \mathcal{B}_{i}(k;\bm{x})}\left(\mu(\ell_{i,j}^{*}) - \mu(\ell_{i,j})\right) v_{i,j} \\
~\leq~ & (1- \beta) \cdot \left( \sum_{j\in \mathcal{B}_{i}(k;\bm{x})}\mu(\ell_{i,j})v_{i,j} + \sum_{j\notin \mathcal{B}_{i}(k;\bm{x})}\mu(\ell_{{k},j})v_{{k},j}+\sum_{j\in \mathcal{B}_{i}(k;\bm{x})}\mu(\ell_{{k},j})\left(v_{{k},j} - v_{i,j}\right)\right)  \\
~=~ & (1- \beta) \cdot \left( \sum_{j\in \mathcal{B}_{i}(k;\bm{x})}\mu(\ell_{i,j})v_{i,j} + \sum_{j\in[M]}  \mu(\ell_{{k},j})v_{{k},j} -\sum_{j\in \mathcal{B}_{i}(k;\bm{x})}\mu(\ell_{{k},j}) v_{i,j}\right)  \\
~\overset{(a)}{\leq}~ & (1- \beta) \cdot \left( \sum_{j\in \mathcal{B}_{i}(k;\bm{x})}\mu(\ell_{i,j})v_{i,j} + \sum_{j\in[M]}  \mu(\ell_{{k},j})v_{{k},j} -\sum_{j\in \mathcal{B}_{i}(k;\bm{x})}\mu(\ell_{i,j}^{*}) v_{i,j}\right)  \,.\\
\Longrightarrow &  \sum_{j\in \mathcal{B}_{i}(k;\bm{x})}\left(\mu(\ell_{i,j}^{*}) - \mu(\ell_{i,j})\right) v_{i,j} ~\leq ~ \frac{1-\beta}{1-\frac{1}{\delta}}\sum_{j\in[M]}  \mu(\ell_{{k},j})v_{{k},j} 
\end{aligned}
\end{align}
where (a) follows from $\mu(\ell_{i,j}^{*}) \leq \mu(\ell_{{k},j})$ due to the fact that $\ell_{{k},j} < \ell_{i,j}$ for any $k\in \mathcal{C}$ and $j\in \mathcal{B}_{i}(k;\bm{x})$.

Summing the above over all $k \in \mathcal{C}$, and following the same arguments as in Eq.\eqref{eq:vcgboundloss} of the proof of Theorem \ref{thm:fairnessbound}, we have

\begin{align}
\begin{aligned}
  \loss_{i}(\bm{\outcome}) ~=~& \sum_{j \in \mathcal{L}_{i}(\bm{\outcome})}\left(\mu(\ell_{i,j}^{*}) - \mu(\ell_{i,j})\right) v_{i,j}\\
  ~\leq~& \sum_{k \in \mathcal{C}}\sum_{j\in \mathcal{B}_{i}(k;\bm{x})} \left(\mu(\ell_{i,j}^{*}) - \mu(\ell_{i,j})\right) v_{i,j}\\
~\leq~& \frac{1-\beta}{1-\frac{1}{\delta}}\sum_{k \in \mathcal{C}}\sum_{j\in[M]}  \mu(\ell_{{k},j})v_{{k},j} \\
~=~& \frac{1-\beta}{1-\frac{1}{\delta}}\sum_{k \in \mathcal{C}}  \well_{k}(\bm{\outcome}) \\
~\leq~& \frac{1-\beta}{1-\frac{1}{\delta}} \well_{-i}(\bm{\outcome}) \\
~\leq~& \frac{1-\beta}{1-\frac{1}{\delta}} \left(\opt_{-i} +  \loss_{i}(\bm{\outcome}) \right) \,.
  \end{aligned}
\end{align}
Rearranging we get $ \loss_{i}(\bm{\outcome}) \leq \frac{1-\beta}{\beta-\frac{1}{\delta}}\opt_{-i} = \frac{1-\beta}{\beta - \frac{\Delta}{2\Delta-1}}\opt_{-i}$. Finally, applying Proposition \ref{prop:boundloss} w.r.t. upper bound of $\loss_{i}(\bm{x})$ and using the fact that the competing bid profile is arbitrary, we obtain the desired welfare guarantee lower bound.
\halmos

\end{APPENDICES}

\end{document}